\documentclass[10pt,a4paper]{article}
\usepackage{citesort,cite}

\usepackage{graphicx,amsmath,amssymb,amsthm,tikz,pgfplots,bm}

\providecommand*{\mrm}[1]{\mathrm{#1}}

\newcommand{\cf}{{\it cf.\/}\ }

\newcommand{\ie}{{\it i.e.\/}, }
\newcommand{\eg}{{\it e.g.\/}, }

\newcommand{\set}[1]{\mathrm{#1}}

\newcommand{\RR}{\mathbb{R}}

\newcommand{\LL}{\mathcal{L}}

\newcommand{\iu}{\mathrm{i}}

\newcommand{\Laplace}{\Delta}

\newcommand{\lexp}[1]{\mathrm{e}^{#1}}
\newcommand{\sbj}{\mathrm{j}}

\newcommand{\RE}{\mathop{\set{Re}}}
\newcommand{\IM}{\mathop{\set{Im}}}

\DeclareMathOperator{\Eint}{E}
\DeclareMathOperator{\Kint}{K}
\DeclareMathOperator{\Fint}{F}

\newcommand{\diff}{\mathop{\mathrm{\mathstrut{d}}}\!}

\newcommand{\Oh}{\mathcal{O}}

\newcommand{\Div}{\nabla\cdot}
\newcommand{\Rot}{\nabla\times}

\newcommand{\dotp}[2]{\langle #1,#2\rangle}

\newcommand{\eps}{\varepsilon}

\newcommand{\vJ}{\boldsymbol{J}}
\newcommand{\vA}{\boldsymbol{A}}
\newcommand{\vB}{\boldsymbol{B}}

\newcommand{\vr}{\boldsymbol{r}}
\newcommand{\vR}{\boldsymbol{R}}

\newcommand{\he}{\hat{\bm{e}}}
\newcommand{\hx}{\hat{\bm{x}}}

\newcommand{\hz}{\hat{\bm{z}}}
\newcommand{\hk}{\hat{\bm{k}}}
\newcommand{\hn}{\hat{\bm{n}}}
\newcommand{\hr}{\hat{\bm{r}}}

\newcommand{\vE}{\boldsymbol{E}}
\newcommand{\vH}{\boldsymbol{H}}
\newcommand{\vD}{\boldsymbol{D}}

\newcommand{\KK}{\mathcal{K}}

\newcommand{\QQ}{\mathcal{Q}}

\newcommand{\ju}{\mathrm{j}}
\newcommand{\Eo}{\bm{F}_{\mathrm{E}}}
\newcommand{\Ho}{\bm{F}_{\mathrm{H}}}
\newcommand{\hR}{\hat{\boldsymbol{R}}}

\newcommand{\hatm}{\hat{\boldsymbol{m}}}
\newcommand{\hp}{\hat{\boldsymbol{p}}}
\newcommand{\vp}{\boldsymbol{p}}
\newcommand{\vm}{\boldsymbol{m}}
\newcommand{\vpe}{\boldsymbol{p}_{\mrm{e}}}
\newcommand{\vpm}{\boldsymbol{p}_{\mrm{m}}}
\newcommand{\vme}{\boldsymbol{m}_{\mrm{e}}}
\newcommand{\vmm}{\boldsymbol{m}_{\mrm{m}}}

\newcommand{\roe}{\rho_{\mrm{e}}}
\newcommand{\rom}{\rho_{\mrm{m}}}

\newcommand{\Rr}{\RR_r^3}

\newcommand{\Wez}{\We^{(0)}}

\newcommand{\Wmz}{\Wm^{(0)}}

\newcommand{\pPz}{P^{(0)}}
\newcommand{\rPz}{P_{\mrm{rad}}^{(0)}}
\newcommand{\rP}{P_{\mrm{rad}}}
\newcommand{\Pe}{P_{\mrm{e}}}
\newcommand{\Pm}{P_{\mrm{m}}}

\newcommand{\Js}{\vJ_{\mrm{s}}}
\newcommand{\rs}{\rho_{\mrm{s}}}
\newcommand{\rms}{\rho_{\mrm{ms}}}

\newcommand{\lde}{\tilde{\pi}_{\mrm{e}}}
\newcommand{\tde}{\tilde{\boldsymbol{\pi}}_{\mrm{e}}}
\newcommand{\hde}{\hat{\boldsymbol{\pi}}_{\mrm{e}}}
\newcommand{\de}{\boldsymbol{\pi}_{\mrm{e}}}
\newcommand{\dm}{\boldsymbol{\pi}_{\mrm{m}}}
\newcommand{\vJv}{\mathbf{J}_{\mrm{v}}}
\newcommand{\vJm}{\boldsymbol{J}_{\mrm{m}}}
\newcommand{\vJe}{\boldsymbol{J}_{\mrm{e}}}
\newcommand{\Jmo}{\boldsymbol{J}_{\mrm{m},1}}
\newcommand{\Jeo}{\boldsymbol{J}_{\mrm{e},1}}
\newcommand{\Jmt}{\boldsymbol{J}_{\mrm{m},2}}
\newcommand{\Jet}{\boldsymbol{J}_{\mrm{e},2}}
\newcommand{\Jem}{\boldsymbol{J}_{\mrm{e},\mrm{m}}}

\newcommand{\We}{W_{\mrm{e}}}
\newcommand{\Wm}{W_{\mrm{m}}}
\newcommand{\Wem}{W_{\mrm{em}}}

\newcommand{\Qe}{Q_{\mrm{e}}}
\newcommand{\Qm}{Q_{\mrm{m}}}

\newcommand{\gae}{\boldsymbol{\gamma}_{\mrm{e}}}
\newcommand{\gam}{\boldsymbol{\gamma}_{\mrm{m}}}
\newcommand{\pim}{\phi_{\mrm{m}}}

\newcommand{\hh}{\hat{\boldsymbol{h}}}
\newcommand{\imp}{\eta}

\newcommand{\Le}{\mathcal{L}_\mrm{e}}
\newcommand{\Lm}{\mathcal{L}_\mrm{m}}
\newcommand{\Lem}{\mathcal{L}_\mrm{em}}

\newcommand{\Jve}{\vJe}
\newcommand{\Jvm}{\vJm}

\newcommand{\psiv}{\boldsymbol{\psi}}
\newcommand{\Xmm}{\mathbf{X}\Ma}
\newcommand{\Xme}{\mathbf{X}\El}
\newcommand{\Ma}{_\mrm{m}}
\newcommand{\El}{_\mrm{e}}

\newcommand{\herm}{\text{H}}
\newcommand{\Fm}{\mathbf{E}_\infty}

\newcommand{\minimize}{\mathop{\mrm{minimize}}}

\newcommand{\maximize}{\mathop{\mrm{maximize}}}

\newcommand{\subto}{\mrm{subject\ to\ }}

\title{Stored energies in electric and magnetic current densities for small antennas}
\author{B.L.G.~Jonsson$^{1}$ and Mats Gustafsson$^{2}$\\
{\small $^1$KTH Royal Institute of Technology, School of Electric Engineering, Teknikringen 33, SE-100 44 Stockholm, Sweden}\\
{\small $^2$Department of Electrical and Information Technology, Lund University, Box 118, SE-221 00 Lund, Sweden}}
\title{Stored energies in electric and magnetic current densities for small antennas}
\begin{document}


\maketitle

\begin{abstract}
Electric and magnetic currents are essential to describe electromagnetic stored energy, as well as the associated quantities of antenna Q and the partial directivity to antenna Q-ratio, $D/Q$, for general structures. The upper bound of previous $D/Q$-results for antennas modeled by electric currents is accurate enough to be predictive, this motivates us here to extend the analysis to include magnetic currents. In the present paper we investigate antenna Q bounds and $D/Q$-bounds for the combination of electric- and magnetic-currents, in the limit of electrically small antennas. This investigation is both analytical and numerical, and we illustrate how the bounds depend on the shape of the antenna. We show that the antenna Q can be associated with the largest eigenvalue of certain combinations of the electric and magnetic polarizability tensors. The results are a fully compatible extension of the electric only currents, which come as a special case. The here proposed method for antenna Q provides the minimum $Q$-value, and it also yields families of minimizers for optimal electric and magnetic currents that can lend insight into the antenna design. 
\end{abstract}


\section{Introduction}

Time harmonic electromagnetic radiating systems do not in general have a finite total energy associated with them. This is well known since the radiated electric and magnetic fields decay as $r^{-1}$ and the corresponding energy density hence decay as $r^{-2}$, which is not an integrable quantity for exterior unbounded regions. This non-integrability differs from the singularities of the electromagnetic energy for charged particles, see \eg~\cite{Dirac1938,Yaremko2003}, where the challenge is the finite mass of particles in coupling Maxwell's equation to the dynamics of the charged particles. 

To consistently extract a finite stored energy from the energy densities associated with classical time-harmonic energy has been investigated in~\cite{Chu1948,Harrington1961,Collin+Rothschild1964,Foltz+McLean1999,Sten+etal2001,Thal2006,Yaghjian+Best2005}.
These stored energies have been based on spherical (and spheroidal) modes, circuit equivalents and on the input impedance for small antennas. 
In 2010 Vandenbosh~\cite{Vandenbosch2010} proposed a current-density approach to stored energies also applicable to larger antennas. This approach has generated new interest in electromagnetic stored energy that is explored in~\cite{Vandenbosch2011,Gustafsson+etal2012a,Capek+etal2012,Gustafsson+Jonsson2013,Gustafsson+Nordebo2013,Vandenbosch2013}. This `stored energy' is similar to the results of Collin and Rothschild~\cite{Collin+Rothschild1964} and it also has similarities with the stored energies proposed in~\cite{Geyi2003,Capek+etal2013}. The generalization in~\cite{Yaghjian+etal2013} and in the present paper includes electric and magnetic current-densities for arbitrary shapes. Antennas embedded in lossy or dispersive material has been considered in~\cite{Gustafsson+etal2014}.

The drive to find a well-defined stored energy stems partly from that it is closely related to the antenna quality factor Q.  Lower bounds on antenna Q is directly related to the electric size of the antenna, and indirectly to the maximal matching bandwidth that can be obtained. The relation between antenna Q and bandwidth is not trivial, for a discussion and examples see \eg~\cite{Yaghjian+Best2005,Gustafsson+Nordebo2006,Gustafsson+Jonsson2013,Gustafsson+Jonsson2014}. 
An alternative method to derive bandwidth bounds is sum-rules, see \eg~\cite{Fano1950,Rozanov2000,Yaghjian+Best2005,Gustafsson+etal2007a,Jonsson+etal2013,Jonsson2014}. The approach given here, is related to~\cite{Yaghjian+Best2005,Gustafsson+etal2007a,Gustafsson+etal2012a,Gustafsson+Jonsson2013,Yaghjian+etal2013}.  In the present paper, we show how the electric and magnetic polarizabilities~\cite{Schiffer+etal1949,Kleinman+Senior1986,Sihvola+Yla+Jarvenpaa+Avelin2004} are directly related to lower bounds on antenna Q for small antennas. 
The investigation is based on the asymptotic behavior of stored energy in the electrically small case for both electric and magnetic currents. This result is an extension of the stored energies in~\cite{Vandenbosch2010} and their connection to both antenna Q and partial directivity over antenna Q. That scattering properties are related to the polarizabilities are known, see e.g.~\cite{Rayleigh1871,Bethe1944}, but that polarizability tensors appear directly as the essential factor in antenna Q-estimates is a recent result~\cite{Gustafsson+etal2007a,Yaghjian+Stuart2010,Jonsson+Gustafsson2013a,Yaghjian+etal2013}. 

The current-representation approach to stored energy enables the maximal partial directivity over antenna Q problem to be reduced to a convex optimization problem~\cite{Gustafsson+Nordebo2013}. It also enables us to consider fundamental limitations for arbitrary geometries. Convex optimization problems are efficiently solvable~\cite{Boyd+Vandenberghe2004}. From a user perspective it can be compared with solving a matrix equation. To numerically find the physical bounds on antenna Q or partial directivity over antenna Q, $D/Q$ is here reduced to tractable problems, solvable with common electromagnetic tools. In the present paper we illustrate how this can be applied to a range of shapes, both numerically and analytically. The here considered minimization problems investigate how different current and charge density combinations yield different lower bounds on antenna Q. For the electrical dipole problem we show that the minimizing currents result in Q and $D/Q$ that agree with~\cite{Gustafsson+etal2007a,Yaghjian+Stuart2010,Thal2006}. For the case of a generalized electric dipole with both electric charges and magnetic current-densities as sources our result agrees with the sphere in~\cite{Chu1948}. When we allow dual-modes, \ie both electrically and magnetically radiation dipoles, we find that the result agree with~\cite{Jonsson+Gustafsson2013a,Yaghjian+etal2013}. The framework here easily account for all these different cases with a generic approach.  
Another result of our method is that we can show that small antennas have a family of current-densities that realize the associated optimal antenna Q for a given shape~\cite{Gustafsson+etal2012a}. 

The present paper is based on the stored energies for both electric and magnetic current densities~\cite{Jonsson+Gustafsson2013a}. Another approach to these energies and associated bounds are given in~\cite{Yaghjian+etal2013}. We investigate the small antenna limit and illustrate how antenna Q and related optimization problems behave for electric and magnetic currents for a range of antenna shapes. These results are based on the leading order terms of the stored energies as the electric size of the domain approach zero. One of the advantages here is that the bounds on $Q$ and $D/Q$ are known once the polarizability tensors are determined for a given shape. We use this knowledge to sweep shape-parameters to illustrate how $Q$ and $D/Q$ depend on the shape of antenna.  
Analytical expressions for the electrically small case provide physical insight into limiting factors for $Q$ and $D/Q$. These more general results are shown to reduce to the analytically known cases in~\cite{Yaghjian+Stuart2010,Gustafsson+etal2009a,Yaghjian+etal2013}. 

In Section~\ref{2}, we recall the definitions of key antenna and energy quantities. Using an asymptotic expansion of the electric and magnetic currents in Section~\ref{3} we give the explicit leading order current-density representation of the radiated power, stored energies, and the radiation intensities. Analytical and numerical examples for antenna Q and $D/Q$ under different constraints are given in Section~\ref{5}. In Section~\ref{4}, we formulate the problem as a convex optimization problem, and determine $Q$ for some shapes. Conclusions and appendices end the paper. 

\section{Antenna Q and partial directivity}\label{2}
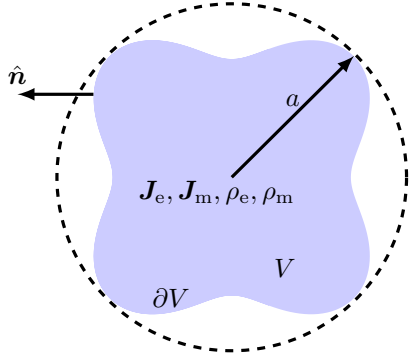
\begin{figure}
\centering
\begin{tikzpicture}[scale=1,very thick]
\draw[fill,color=blue!20,samples=200,smooth] plot (canvas polar
cs:angle=\x r,radius={54-10*cos(4*\x r)});
\draw[dashed] (0,0) circle (23mm);
\draw (-.2,-.2) node {$\vJe,\vJm,\roe,\rom$};
\draw[-latex] (0,0) -- (1.61,1.61) node [midway,above] {$a$};
\draw[-latex] (-1.82,1.1) -- +(-1,0) node [above] {$\hn$};
\draw (.7,-1.2) node {$V$} (-.8,-1.6) node {$\partial V$};
\end{tikzpicture}
\caption{The figure illustrates the joint support, $V$, of the current densities, enclosed within a sphere of radius $a$, and with a normal $\hn$ at the boundary $\partial V$ of $V$.}\label{fig1}
\end{figure}

Let $V\subset \RR^3$ be the joint bounded support of the electric and magnetic current densities $\vJe$ and $\vJm$ respectively, see Figure~\ref{fig1}. The support $V$ is here assumed to be bounded and connected. Through the continuity equations we define the associated electric and magnetic charge densities $\roe$ and $\rom$. The time-harmonic Maxwell's equations with electric and magnetic current densities in free-space take the form
\begin{align}
\label{me1}\nabla \times \vE +\ju \imp k \vH &= - \vJm,&& 
\nabla \cdot \vE = \frac{\roe}{\eps} = \frac{-\imp}{\ju k}\nabla\cdot \vJe ,
\\
\label{me2}\nabla \times \vH - \frac{\ju k}{\imp} \vE & = \vJe , &&
 \ \
\nabla\cdot \vH = \frac{\rom}{\mu} = \frac{-1}{\ju k \imp}\nabla\cdot \vJm,
\end{align}
where we use the time convention $\lexp{\ju\omega t}$, which is suppressed. 
In this paper, we let $\eps=\eps_0$, $\mu=\mu_0$ and $\imp=\imp_0=\sqrt{\mu/\eps}$ be the free space permittivity, permeability and impedance, respectively. $\vE$ is the electric field and $\vH$ is the magnetic field. The dispersion relation between the wave number, $k$, and the angular frequency, $\omega$, is $k=\omega\sqrt{\eps\mu}$ and $t$ is time. 

The field energy densities are $\eps|\vE|^2/4$ and $\mu|\vH|^2/4$. Here we are interested in stored electric $\We$ and magnetic $\Wm$ energies, which are more challenging to define. We follow the definition of~\cite{Collin+Rothschild1964,Fante1969,Geyi2003,Yaghjian+Best2005,Vandenbosch2010,Gustafsson+Jonsson2013,Gustafsson+Jonsson2014} and define the stored electric and magnetic energies as 
\begin{equation}\label{WEM}
\We = \frac{\eps}{4}\int_{\Rr} |\vE(\vr)|^2 - \frac{|\Eo(\hr)|^2}{r^2} \diff V,\ \ \ 
\Wm = \frac{\mu}{4}\int_{\Rr} |\vH(\vr)|^2 - \frac{|\Ho(\hr)|^2}{r^2} \diff V,
\end{equation}
where $\Eo,\Ho$ are the far-fields, \ie $\vE \rightarrow \Eo \frac{\lexp{-\ju k r}}{r}$ as $r\rightarrow \infty$ and $\eta\Ho=\hr\times \Eo$. Let $\vr$ denote a vector in $\RR^3$, with length $r=|\vr|$ and corresponding unit vector $\hr=\vr/r$. Here $\int_{\Rr}$ is an abbreviation of the limit $\lim_{r_0\rightarrow \infty} \int_{|\vr|<r_0}$. Note that the expressions~\eqref{WEM} can for certain antennas become coordinate dependent, and for large structures \eqref{WEM} may become negative~\cite{Gustafsson+etal2012a}, these artifacts do not appear in the small electrical limit, as shown later in this paper as all obtained minimal antenna Q are non-negative, see \eg Section~\ref{5}. 

Given these stored energies, we define the two main antenna parameters that appear in the physical bounds. The antenna quality factor: $Q=\max(\Qe,\Qm,0)$ where
\begin{equation}\label{qeqm}
\Qe=\frac{2\omega \We}{\rP}, \ \text{and} \ \Qm=\frac{2\omega \Wm}{\rP}.
\end{equation}
Here, $\rP$ is the radiated power of the system described by~\eqref{me1}-\eqref{me2}. Defined as 
\begin{equation}\label{rad_op}
\rP=\frac{1}{2\imp}\int_{\Omega} |\Eo(\hr)|^2\diff \Omega,
\end{equation}
where $\Omega$ is the unit sphere in $\RR^3$. 

The partial directivity $D(\hk,\he)$ in the direction $\hk$ from an antenna with polarization $\he$, is \cite{IEEEantennaterms1993}
\begin{equation}
D(\hk,\he) =  4\pi \frac{P(\hk,\he)}{\rP},
\end{equation}
where $P(\hk,\he)$ is the partial radiation intensity $|\he\cdot\Eo|^2/(2\eta)$. The other main antenna parameter here is the partial directivity over antenna Q, $D/Q$, which with the above notation is
\begin{equation}
\frac{D(\hk,\he)}{Q} = \frac{2\pi P(\hk,\he)}{\omega \max(\We,\Wm,0)}.
\end{equation}

The goal here is to optimize and investigate $Q$ and $D/Q$ in terms of the electric and magnetic current densities, in the small antenna limit. We hence express these quantities in terms of the current densities, see~\ref{app:result}. While these calculations are straight forward, they are also rather lengthy, see \eg~\cite{Vandenbosch2010} for a similar effort, see also~\cite{Jonsson+Gustafsson2013a,Jonsson+Gustafsson2013b}. 
Substantial simplification is obtained in these derivations for the case of electrically small antennas which is illustrated in the next section. The leading order term of the stored energies, for small $k$ is given by
\begin{equation}\label{we}
\We = \frac{\mu}{4k}\IM \big[\dotp{\vJe}{\Le\vJe}+\frac{1}{\eta^2}\dotp{\vJm}{\Lm\vJm} \big] + \Oh(k)
\end{equation}
and
\begin{equation}\label{wm}
\Wm = \frac{\mu}{4k}\IM \big[\dotp{\vJe}{\Lm\vJe}+\frac{1}{\eta^2}\dotp{\vJm}{\Lm\vJm} \big] + \Oh(k) . 
\end{equation}
Note that these stored energies are symmetric in the current densities and a natural extension of the electric only current-case, $\vJm=0$. 
Above we use the ordo notation $\Oh(k)$ to indicate that the next order term is bounded by $Ck$, for some constant $C$ as $k\rightarrow 0$. The associated operators in~\eqref{we} and \eqref{wm} are
\begin{align}\label{Le}
\dotp{\vJ}{\Le\vJ} &= \frac{-1}{\ju k}\int_V\int_V \nabla_1\cdot \vJ(\vr_1)\nabla_2\cdot \vJ^*(\vr_2)G(\vr_1-\vr_2)\diff V_1\diff V_2,\\\label{Lm}
\dotp{\vJ}{\Lm\vJ} &= \ju k \int_V\int_V  \vJ(\vr_1)\cdot \vJ^*(\vr_2)G(\vr_1-\vr_2)\diff V_1\diff V_2.
\end{align}
The operators are similar to the Electric Field Integral Equation, EFIE, operators $\LL=\Le-\Lm$, when the currents are on a surface of an object, and for such currents there is a range of implementations in the standard method-of-moment codes. 
Here and below we occasionally use the notation `current', in place of `current density', to shorten the notation. 
The kernel $G(\vr)$ is the Green's function, $\lexp{-\ju kr}/(4\pi r)$ and $*$ indicates the complex conjugate, see also Figure~\ref{fig1}. 

The radiation intensity, $P(\hk)$ in the direction $\hk$, have a representation in terms of the current densities~\cite{Jonsson+Gustafsson2013a}:
\begin{multline}\label{intP}
P(\hk)= \frac{\eta k^2}{32\pi^2}\Big[ \big|\int_V (\he^*\cdot \vJe(\vr_1) + \frac{1}{\eta}\hk\times \he^* \cdot \vJm(\vr_1))\lexp{\ju k \hr\cdot\vr_1} \diff V_1 \big|^2  + \\ \big|\int_V (\hh^*\cdot \vJe(\vr_1) + \frac{1}{\eta}\hk\times \hh^*\cdot \vJm(\vr_1) )\lexp{\ju k \hr\cdot\vr_1} \diff V_1\big|^2 \Big]  =P(\hk,\he) + P(\hk,\hh),
\end{multline}
where we use that $\hk,\he,\hh$ is an orthogonal triplet with $\hk\times \he=\hh$. We recognize the partial radiation intensity $P(\hk,\he)$ for the polarization $\he$.  For electric currents only, \ie $\vJm=0$, these expression agree with \eg~\cite{Vandenbosch2010,Gustafsson+etal2012a}.

To find the total radiated power $\rP$, in terms of its current-density representation we can integrate~\eqref{intP} over the unit sphere. A more direct route to $\rP$ is based on~\eqref{rad_op} and the observation that the electric far-field, $\Eo$, have the representation
\begin{equation}\label{Eo}
\Eo(\hr) = \frac{\ju\imp k }{4\pi} \hr\times \int_{V} \Big[\hr\times \vJe(\vr_1) + \frac{1}{\eta} \vJm(\vr_1)\Big]\lexp{\ju k \hr\cdot \vr_1}\diff V_1.
\end{equation}
Somewhat lengthy calculations~\cite{Jonsson+Gustafsson2013b,Jonsson+Gustafsson2013a} show that the corresponding quadratic form in terms of the currents are 
\begin{equation}\label{oprP}
\rP = \frac{\eta}{2}\RE \dotp{\vJe}{\LL\vJe}  + \frac{1}{2\eta}\RE\dotp{\vJm}{\LL\vJm} - \IM \dotp{\vJe}{\KK_1\vJm},
\end{equation}
where $\KK_1$ is the operator defined by
\begin{equation}
\dotp{\vJe}{\KK_1\vJm}= \frac{k^2}{4\pi}\int_V\int_V \vJe^*(\vr_1)\cdot \hR\times \vJm(\vr_2)\sbj_1(kR)\diff V_1\diff V_2.
\end{equation}
Here $\vR=\vr_1-\vr_2$, $R=|\vR|$, $\hR=\vR/R$ and $\sbj_n(x)$ is the spherical Bessel function of order $n$~\cite{Abramowitz+Stegun1970}. 

The small electrical size limit simplify the above energy and power related expressions $\We,\Wm,P(\hk)$ and $\rP$ and subsequently $Q$ and $D/Q$. We utilize that the radius, $a$, of the enclosing sphere, is electrically small, \ie that $ka$ is small enough to motivate that we discard higher order terms. To expand the above quantities in terms of small $ka$ we assume that the currents have the asymptotic behavior:
\begin{align}\label{exp}
\vJe = \vJe^{(0)} + k \vJe^{(1)} + \Oh(k^2) \ \text{with}\ \Div\vJe^{(0)} = 0,  \\ 
\vJm = \vJm^{(0)} + k \vJm^{(1)} + \Oh(k^2) \ \text{with} \ \Div\vJm^{(0)} = 0.\label{exp2}
\end{align}
This assumption is consistent with the continuity equations for the electric and magnetic current densities. Note that $\vJe^{(0)}$, $\vJm^{(0)}$, $\vJm^{(1)}$ and $\vJm^{(1)}$ are all $k$-independent and the two latter correspond to a lowest order static charge through the continuity equation.  

\section{Electrically small volume approximation}~\label{3}

We apply the small $ka$ approximation and~\eqref{exp}, \eqref{exp2} to the partial radiation intensity and the far-field $\Eo$ in the form of~\eqref{Eo}. We first note that 
\begin{equation}\label{cur}
\int_V \vJ\lexp{\ju k \hk\cdot \vr} \diff V = 
\int_V \vJ^{(0)} + k\vJ^{(1)} + \ju k (\hk\cdot\vr)\vJ^{(0)} + \Oh(k^2)\diff V = -\ju k \int_V \ju \vJ^{(1)} +\frac{1}{2}\hk\times(\vr\times \vJ^{(0)}) \diff V+\Oh(k^2),
\end{equation}
where we have used that~\cite[p432]{Stratton1941}:
\begin{equation}\label{zero}
\int_{V} \Jem^{(n)}\diff V = \left\{\begin{array}{ll} 0, & n=0, \\-\int_{V} \vr \Div\Jem^{(n)} \diff V, & n\neq 0 ,\end{array}\right.
\end{equation}
and \cite[p433]{Stratton1941},~\cite[p127]{Kristensson1999a}
\begin{equation}\label{zeroone}
\int_V (\hk\cdot \vr)\Jem^{(0)}\diff V=\frac{-1}{2}\hk\times \int_V \vr\times \Jem^{(0)}\diff V,
\end{equation} 
since $\Div\Jem^{(0)}=0$. Here $\Jem^{(n)}$, indicate that the expression is valid for $\vJe^{(n)}$ and $\vJm^{(n)}$, $n=0,1$. It follows that the partial radiation intensity~\eqref{intP}, for a wave with polarization $\he$ and propagating in direction $\hk$ is $P(\hk,\he)=\pPz(\hk,\he)+\Oh(k^5)$, where $\pPz$ reduces to 
\begin{multline}\label{ipsmall}
\pPz(\hk,\he)=  
 \frac{\eta k^4}{32\pi^2} \big|\int_V \he^*\cdot (\ju  \vJe^{(1)}  + \frac{1}{2\eta}\vr\times\vJm^{(0)} ) +
\hk\times \he^*\cdot(\frac{\ju}{\eta}\vJm^{(1)}-\frac{1}{2}\vr\times \vJe^{(0)})   \diff V \big|^2\\= \frac{\eta k^4}{32\pi^2} \big| \he^*\cdot \de + \hk\times \he^*\cdot \dm\big|^2.
\end{multline}
Here we used that the triplet $\hk,\he^*,\hh^*$ forms an orthogonal basis system. The   
\begin{equation}\label{genpm}
\de=\int_V \ju\vJe^{(1)}+\frac{1}{2\eta}\vr\times\vJm^{(0)}\diff V\ \text{and}\
\dm=\int_V \frac{\ju}{\eta}\vJm^{(1)}-\frac{1}{2}\vr\times\vJe^{(0)}\diff V
\end{equation} 
terms are generalized dipole-moments that account for both the electric and magnetic dipole radiating fields, respectively. 

To find the total radiated power in~\eqref{rad_op} we start with inserting the expansion \eqref{cur} into the far-field~\eqref{Eo} to find the small $ka$ approximation of the far-field: 
\begin{equation}\label{eoo}
\Eo(\hk) = \frac{\eta k^2}{4\pi}\hk \times \int_V \hk\times (\ju \vJe^{(1)}+\frac{1}{2\eta}\vr\times \vJm^{(0)}) + (\frac{\ju}{\eta}\vJm^{(1)}-\frac{1}{2} \vr\times \vJe^{(0)}) \diff V + \Oh(k^3).
\end{equation}
We insert~\eqref{eoo} into the expression for the total radiated power \eqref{rad_op}, to find that $\rP=\rPz+\Oh(k^5)$ where
\begin{multline}\label{smallrP}
\rPz=\frac{\eta k^4}{32\pi^2} \int_{\Omega} |\de|^2 - |\hk\cdot \de|^2 + |\dm|^2 - |\hk\cdot \dm|^2 - 2\hk\cdot \RE (\dm\times \de^*) \diff \Omega  = 
\frac{\eta k^4}{12\pi} (|\de|^2+|\dm|^2) \\ = 
\frac{\eta k^4}{12\pi} \Big\{
\Big|\int_V \ju \vJe^{(1)}+\frac{1}{2\eta}\vr\times\vJm^{(0)}\diff V\Big|^2 +
\Big|\int_V \frac{\ju}{\eta} \vJm^{(1)}-\frac{1}{2}\vr\times\vJe^{(0)}\diff V\Big|^2 \Big\}=\Pe+\Pm.
\end{multline}
Here we used the integration over the unit sphere $\Omega$ of the angular variables in $\hk$ to find the relations $\int_{\Omega} \hk\diff\Omega=0$ and $\int_{\Omega} |\hk\cdot\de|^2\diff \Omega = \frac{4\pi}{3}|\de|^2$. 
The radiated power consists of two types of terms: terms that radiate as electric dipoles with power $\Pe$ and the second part that radiates as magnetic dipoles with power $\Pm$. An alternative derivation to calculate $\rP$ in the small volume limit is to start from~\eqref{oprP}, see~\ref{app:long}. 
The power in terms of the dipole-moments can alternatively be expressed as
\begin{equation}
\rP =  
 \frac{k^4}{12 \pi \sqrt{\eps\mu}}\Big[ \big|\frac{1}{\sqrt{\eps}}\vpe - \sqrt{\eps}\vmm\big|^2  
   + \big|\frac{1}{\sqrt{\mu}}\vpm +\sqrt{\mu}\vme\big|^2 \Big] 
+\Oh(k^5) ,
\end{equation}
where $\ju c \vpe=\int_V\vJe^{(1)}\diff V$ and $\vme=\frac{1}{2}\int_V\vr\times \vJe^{(0)}\diff V$ and analogously for the magnetic currents and moments with subscript m, \ie $\vmm$. Here $c=1/\sqrt{\eps\mu}$ is the speed of light.

A check that the above expressions agree with what is known for small antennas that radiate as dipoles is obtained by comparing the maximal partial directivity, \ie $\pPz(\hk,\he)$ to the total radiation $\rP$. We consider two cases: fixed generalized electric dipole moments and no generalized magnetic dipole moment~\eqref{genpm} \ie $\dm=0$ and $\de\neq 0$ (or vice versa) and fixed non-zero $\dm,\de$:
\begin{equation}\label{PPpure}
\max_{\he}\frac{4\pi \pPz(\hk,\he)}{\rPz} = \frac{3}{2}, \ \text{when}\ \dm=0, \text{ and }
\max_{\he, \he\bot\hk}\frac{4\pi \pPz(\hk,\he)}{\rPz} = \max_{\he,\he\bot\hk}
\frac{3}{2} \frac{|\he^*\cdot( \de - \hk\times \dm)|^2}{|\de|^2+|\dm|^2}\leq 3.
\end{equation}
Stating that a small antenna with electric dipole radiation from a generalized electric dipole moment have directivity $3/2$, but upon adding a magnetic generalized dipole $\dm$ we find that appropriate oriented combinations of $\de$ and $\dm$ can have a directivity of 3, corresponding to a Huygens source see \eg~\cite{Pozar2009}.  

The small electric volume stored energies follow directly from their integral representation~\eqref{we}, we find that $\We =\Wez + \Oh(k)$, where 
\begin{equation}\label{We_asymp}
\Wez = \frac{\mu}{16\pi }
 \int_V  \int_V \bigr[\frac{1}{\imp^2}\vJm^{(0)}(\vr_1)\cdot \vJm^{(0)*}(\vr_2) + (\nabla_1\cdot\vJe^{(1)}(\vr_1))(\nabla_2\cdot\vJe^{(1)*}(\vr_2))\big] \frac{1}{R_{12}}\diff V_1\diff V_2 
\end{equation}
and similarly~\eqref{wm} yields $\Wm = \Wmz + \Oh(k)$, where
\begin{equation}
\Wmz = \frac{\mu}{16\pi}\int_V \int_V \big[ \vJe^{(0)}(\vr_1)\cdot  \vJe^{(0)*}(\vr_2) +\frac{1}{\eta^2} (\nabla_1\cdot\vJm^{(1)}(\vr_1))(\nabla_2\cdot \vJm^{(1)*}(\vr_2) ) \big]\frac{1}{R_{12}} \diff V_1\diff V_2.
\end{equation}

\section{Minimal antenna Q and analytical and numerical illustrations}~\label{5}

One of the goals with the above expressions for antenna Q and $D/Q$ is that they should lend us some insight into antenna design and limitations of $Q$ and $D/Q$. It is reasonable to ask the question of what shapes that give low antenna Q. Similarly we investigate which charge and current densities that gives low antenna Q. Another goal with the expressions is to find easily derived a priori bounds of antenna Q and $D/Q$. Partial answers are given in this section, that extends the relation that a large charge-separation ability in the domain imply a small antenna Q see \eg~\cite{Gustafsson+etal2007a,Gustafsson+etal2009a,Gustafsson+etal2012a,Yaghjian+Stuart2010,Yaghjian+etal2013}. Similarly we may think of a shape with low antenna Q, as a structure that supports a large `current loop area' for a magnetic dipole-moment. One of the new results here is that the generic shape results in~\cite{Gustafsson+etal2012a} for $D/Q$ is extended to lower bounds on antenna Q. 

An often studied case is the electric-dipole case~\cite{Vandenbosch2010,Yaghjian+Stuart2010,Gustafsson+etal2007a,Gustafsson+Nordebo2013,Gustafsson+etal2009a}, here represented by the electric charges only and we illustrate below how an optimization problem is used to determine the minimal $Q$. We continue and show that the method and its associated eigenvalue-problem extend to the more general case of both electric and magnetic currents that radiate as an electrical dipole. Here we also find that the magnetic polarizability enters in the lower bounds on $Q$. A short review of polarizability tensors are given in~\ref{PE}. 

Consider the minimization problem for finding the lower bound on $Q$.
\begin{equation}\label{29}
Q = \minimize_{\roe^{(1)},\rom^{(1)},\vJe^{(0)},\vJm^{(0)}}\frac{2\omega \max\{\Wez(\roe^{(1)},\vJm^{(0)}),\Wmz(\rom^{(1)},\vJe^{(0)}),0\}}{\Pe(\roe^{(1)},\vJm^{(0)})+\Pm(\rom^{(1)},\vJe^{(0)})}  ,
\end{equation}
with the two constraints $\int_V\roe^{(1)}\diff V=0$ and $\int_V\rom^{(1)}\diff V=0$.  Here $\ju \omega \roe^{(1)}=-k\Div\vJe^{(1)}$ and similarly for $\rom^{(1)}$.  
One of the interesting cases in antenna design is when the antenna radiate as an electrical dipole, \ie when $\Pm$ is negligible and $\Wm\leq \We$. Once the optimal $(\Pe,\We)$ is determined we tune the antenna with a tuning circuit to make the antenna resonant, \ie $\Wm=\We$. Thus we start with the optimization problem for a pure $(\We,\Pe)$-case.  The `dual mode' case, where both $\Pe$ and $\Pm$ are comparable is considered in Section~\ref{dual} below. 
Before we consider the general case, let's start with the easier case of an electric dipole when we have only $\roe$, \ie $\vJm^{(0)}=0$. 

\subsection{Antenna Q for an electric dipole, \eg $\Pm=0$}\label{sQe}

Different approaches to lower bounds of this antenna Q case has also been investigated in \eg~\cite{Vandenbosch2010,Yaghjian+Stuart2010,Gustafsson+etal2007a,Gustafsson+Nordebo2013,Gustafsson+etal2009a,Gustafsson+etal2012a,Vandenbosch2011}. However, one of the goals here is to arrive to a generic method that works for different cases of current-density sources, and the first step towards this goal, is to verify that this method indeed gives the previously derived result on the lower bound see \eg~\cite{Thal2006,Gustafsson+etal2007a,Gustafsson+etal2012a,Yaghjian+Stuart2010,Yaghjian+etal2013}. 
The electric dipole is here equivalent with the assumption $\Pm=0$ and $\We\geq \Wm$ which yields that $Q=\Qe$ and that we have an optimization problem that depend only on the electric charge-densities $\roe$. Once the design is determined we can tune the antenna with a tuning circuit to make $\We=\Wm$.  This case is the classical electrical dipole-case.  Let $\roe=\roe^{(1)}$.  The minimization problem~\eqref{29} reduces to:
\begin{equation}\label{mQ}
\Qe= \minimize_{\roe} \frac{2\omega \We^0(\roe)}{\Pe(\roe)}=\frac{6\pi}{k^3}\minimize_{\roe}\frac{\int_V\int_V \frac{\roe^*(\vr_1)\roe(\vr_2)}{4\pi |\vr_1-\vr_2|}\diff V_1\diff V_2}{|\int_V \vr\roe\diff V|^2},
\end{equation}
where we have used~\eqref{zero} to re-write the denominator. This minimization comes with the constraint that no current flows through the surface $\partial V$, \ie $0=\int_{\partial V} \hn\cdot \vJe\diff S=-\ju c \int_V\roe\diff V$, where $c$ is the speed of light. Hence,~\eqref{mQ} is accompanied with the constraint of total zero charge, $\int_V\roe\diff V=0$. 

The associate problem to maximize $D/Q$ in the small electric volume limit for arbitrary $\roe$, see~\cite{Gustafsson+etal2012a} corresponds to:
\begin{equation}\label{dq}
\frac{D}{\Qe} = \maximize_{\roe} \frac{2\pi P^0(\hk,\he)}{\omega \We^0(\roe)} = \frac{k^3}{4\pi} \frac{|\int_V \he^*\cdot \vr\roe(\vr)\diff V|^2}{\int_V\int_V \frac{\roe^*(\vr_1)\roe(\vr_2)}{4\pi |\vr_1-\vr_2|}\diff V_1\diff V_2},
\end{equation}
with the same constraint of a total zero charge, $\int_V \roe\diff V=0$. These two problems are related but the $D/Q$-problem has the simplification in that the integrand in $P^0(\hk,\he)$ see~\eqref{ipsmall}, is scalar-valued and the maximization has a convex optimization formulation see~\cite{Gustafsson+Nordebo2013,Gustafsson+etal2012a}.  

The method that we apply below to~\eqref{mQ} works on both problems \eqref{mQ} and \eqref{dq} and yield the same result as in~\cite{Gustafsson+etal2012a} where it is applied to \eqref{dq}. The final result is similar to the result in~\cite{Yaghjian+Stuart2010,Yaghjian+etal2013}, but obtained with different methods. Note that both \eqref{mQ} and \eqref{dq} remain unchanged under the scaling, $\rho\mapsto\alpha\rho$. Thus the solutions to~\eqref{mQ} are a family of scaling invariant solutions.  We determine the minimum by breaking the scaling-invariance by selecting a particular value of the amplitude of the dipole moment, $p_e$. We rewrite~\eqref{mQ} as the minimization problem as:
	\begin{align}
		  \mathop{\mathrm{minimize}}_{\roe} & \int_V\int_V \frac{\roe^*(\vr_1)\roe(\vr_2)}{4\pi |\vr_1-\vr_2|}\diff V_1\diff V_2, \label{mqea}\\ 
		 \subto &|\int_V \vr \roe(\vr)\diff V|^2 = p_e^2, \label{mqeb}\\
		 & \int_V \roe(\vr) \diff V = 0.\label{mqec}
	\end{align}
This is a classical optimization problem for the Newton-potential. An energy space approach in a similar context is discussed in~\cite{Landkof1972} and an approach that allow for geometries with corners is given in~\cite{Helsing+Perfekt2013}. We note that there may be several minimizers that realize the same minimum, \eg for spheres and shapes with appropriate symmetries~\cite{Sihvola+Yla+Jarvenpaa+Avelin2004}. To explicitly find the minimum, we use the method of Lagrange multipliers see \eg~\cite[\S4.14]{Zeidler1995b} and define the Lagrangian $\QQ$ as 
\begin{equation}\label{eQf}
\QQ(\rho,\rho^*,\lambda_1,\lambda_2) = \int_V\int_V \frac{\rho^*(\vr_1)\rho(\vr_2)}{4\pi |\vr_1-\vr_2|}\diff V_1 \diff V_2 - \lambda_1 (|\int_V \vr\rho\diff V|^2-p_{\mrm{e}}^2) - \lambda_2 \int_V \rho^*\diff V.
\end{equation}
Here $\lambda_1$ and $\lambda_2$ are Lagrange multipliers, and we use the short hand notation $\rho=\roe$.
Variation of $\QQ$ with respect to $\lambda_1$ and $\lambda_2$ gives the two constraints above. 
Taking the variation of $\QQ$ with respect to $\rho^*$, or equivalently, taking a Fr\'echet derivative of $\QQ$ yields the Euler-Lagrange equation for the critical points
\begin{equation}\label{EL0}
\int_V \left( \frac{1}{4\pi |\vr_1-\vr_2|} - \lambda_1 \vr_1\cdot\vr_2\right)\rho(\vr_1) \diff V_1 = \lambda_2, \ \vr_2\in V. 
\end{equation}
Note that this is an integral equation with unknown $\rho$. Accompanied with the constraints we find three equations~\eqref{EL0}, \eqref{mqeb}, and \eqref{mqec} and three unknown $\rho$, $\lambda_1$ and $\lambda_2$. 

Upon multiplying~\eqref{EL0} with $\rho^*$ and integration over $V$, utilizing the zero total charge constraint, we find that $\Qe$ in \eqref{mQ} is equivalent with 
\begin{equation}\label{lambdaQ}
\Qe=\frac{6\pi}{k^3}\min_\rho \lambda_1.
\end{equation}
The unknown Lagrange multiplier, $\lambda_1$, depends implicitly on $\rho$ and $\lambda_2$. The lower bound of the minimization problem~\eqref{mQ} is hence determined by the unknown Lagrange multiplier $\lambda_1$, times a constant. Another property of the solution appears if we apply Laplace operator on~\eqref{EL0}, for $\vr\notin \partial V$ we have that $\rho(\vr)=0$. Thus we reduce~\eqref{EL0} to:
\begin{equation}\label{EL1}
\int_{\partial V} \left(\frac{1}{4\pi |\vr_1-\vr_2|} - \lambda_1 \vr_1\cdot\vr_2\right)\rs(\vr_1) \diff S_1 = \lambda_2, \ \vr_2\in \partial V ,
\end{equation}
where $\rs$ is the surface charge density, \ie we have formally the relation that $\rho\diff V = \rs\diff S$. A similar result for $D/Q$ was shown in~\cite{Gustafsson+etal2012a}.

Using the constraint $|\int_{\partial V} \vr\rs\diff S|=p_e>0$ we re-write the critical equation~\eqref{EL1} into:
\begin{equation}\label{EL2}
\int_{\partial V} \frac{\rs(\vr_1)}{4\pi|\vr_2-\vr_1|} \diff S_1 =  \lambda_1  p_e \hp\cdot \vr_2 +\lambda_2, \ \vr_2\in \partial V 
\end{equation}
for some unknown unit vector $\hp$. 

To solve the equation~\eqref{EL2} we make first a few observations: Any solution $\rs$ of \eqref{EL2} for given right hand-sides, yields an associated potential that solves an electrostatic boundary-value problem \cf~\ref{PE}. Such solutions are restricted in their asymptotic behavior by the electric polarizability tensor $\gae$, which depends only on the shape of $V$. To make this restriction explicit we note that the electric polarizability tensor $\gae$ is defined through~\eqref{pe} and \eqref{Dgae}: $\gae \cdot \he D_0 = \vp$. Comparing~\eqref{pe} with \eqref{EL2}, we see that the generic $D_0\he$ in~\eqref{pe} is here $D_0\he=\lambda_1 p_e \hp$, and the dipole-moment is by definition $\vp=\int_{\partial V} \vr \rs\diff S$. 
Since the electric polarizability tensor $\gae$ is given, once the shape $V$ is known, we thus have a constraint on $(\lambda_1,\hp)$ in order for $\rs$ and its associated potential to comply with the polarizability tensor. The constraint is that: 
\begin{equation}\label{EV1}
\gae\cdot \hp = \frac{1}{\lambda_1}\hp,
\end{equation}
which we recognize as a eigenvalue problem in $(\lambda_1,\hp)$ for $\gae$. Here we have used that $\vp=p_e\hp$. 

We conclude that critical points of~\eqref{mQ} correspond to solutions $(\lambda_1,\hp)$ of the eigenvalue problem \eqref{EV1}. Given such a solution $(\lambda_1,\hp)$ we determine the associated charge-density through~\eqref{EL2} with $(\lambda_1,\hp)$ given as solutions to~\eqref{EV1}. A charge density that solves~\eqref{EL2} is hence the base for the family of current-sources that supports the optimal radiation, which we obtain from the continuity equation. Through the re-writing of the optimal $\Qe$ in~\eqref{lambdaQ} it follows that the largest eigenvalue, $(\gae)_3$ of the polarizability matrix $\gae$ yields the minimum $\Qe$, \ie 
\begin{equation}\label{Qeev}
\Qe= \frac{6\pi}{k^3(\gae)_{3}} .
\end{equation}
We have hence reduced the variational problem of finding the minimum $\Qe$ for the electric dipole to finding eigenvalues of $\gae$. This result have large similarities to~\cite{Yaghjian+Stuart2010,Yaghjian+etal2013}, derived with different methods. We conclude that $Qk^3$ in the small volume size only depend on shape and size as expressed through the electric polarizability. The physical interpretation connects large polarizability eigenvalues to the ability of the structure to separate charge under an external static field in a given direction. The polarizability $\gae$ is associated with the scalar Dirichlet-problem of the Laplace operator, and depend only on the shape of the object~\cite{Schiffer+etal1949}. We note also that $\gae$ is identical with the high-contrast electric polarizability in \eg~\cite{Gustafsson+etal2007a}.

We note that the low-frequency magnetic charge density and electric charge density antenna Q are dual-similar, and hence if we consider a case with either a $\roe$-term or a $\rom$-terms both of these problems result in identical minimization problems with a lower bound on antenna Q given by~\eqref{Qeev}.

To compare with the $D/Q$-problem, we note that the constraint $|\int\he^*\cdot\vr\rs\diff V|=\text{const}$, was in~\cite{Gustafsson+etal2012a} reduced to $\int\he\cdot \vr\rs^*\diff V = \alpha$, yielding the critical equation corresponding to~\eqref{EL2} as
\begin{equation}
\int_{\partial V} \frac{\rs(\vr_1)}{4\pi|\vr_1-\vr_2|} \diff S_1 =  \nu_1\he \cdot \vr_2 + \nu_2, \ \vr_2\in \partial V .
\end{equation}
Similarly to $\Qe$-case above we find that $\nu_1$ is connected to $\gae$ through the relation $\he^*\cdot\gae\cdot\he=\frac{\alpha}{\nu_1}$. The corresponding maximum is $D/Q=\frac{k^3}{4\pi}\he^*\cdot\gae\cdot\he$. We thus see that the two problems are related, but that they describe different optimization problems. The antenna Q lower bound, minimize $Q$ without concern of radiation direction of the antenna, whereas $D/Q$ assume a fixed $\he$ radiation direction though out its optimization. With a-priori knowledge about the optimal radiation direction of the structure or alternatively the principal eigenvalue of $\gae$ associated with a given structure, we select $\he$ in this direction, to find the expected 3/2 difference between $1/\Qe$ and $D/\Qe$. This $D/Q$ result is similar to the sum-rule in~\cite{Gustafsson+etal2007a} for electric sources. With the $\Qe$ result and the observation of principal directions of $\gae$ we see that these three approaches illustrate closely connected results here with a common energy principle method to obtain them. 

To illustrate the result we begin with a sphere:  $\gae=4\pi a^3 \bm{I}$, where $\bm{I}$ is the 3 times 3 unit tensor, and all eigenvalues of $\gae$ are identical. Note that to these degenerate eigenvalues there are three orthogonal eigenvectors, and the corresponding charge densities in~\eqref{EL1} for each a given amplitude of the dipole-moment $p_e$. This degeneracy is due to the geometrical symmetries of the shape. Thus even when we remove the scaling invariance, we may have multiple $\rho$ that yields the same lower bound on $Q$. Note also that for any arbitrary optimizer $\roe=\roe^{(1)}$ that the associated electric current connected to $\roe^{(1)}$, here $\vJe^{(1)}$, \ie $\ju\omega \roe^{(1)}=-\Div\vJe^{(1)}$, has an infinite dimensional subspace that all yield the same $\roe^{(1)}$. It allows a potentially large design-freedom that does not change $\Qe$ in the quasi-static limit. This case is analogous to the case discussed in~\cite{Gustafsson+etal2012a}. 

\begin{figure}[h!]
\centering
\includegraphics{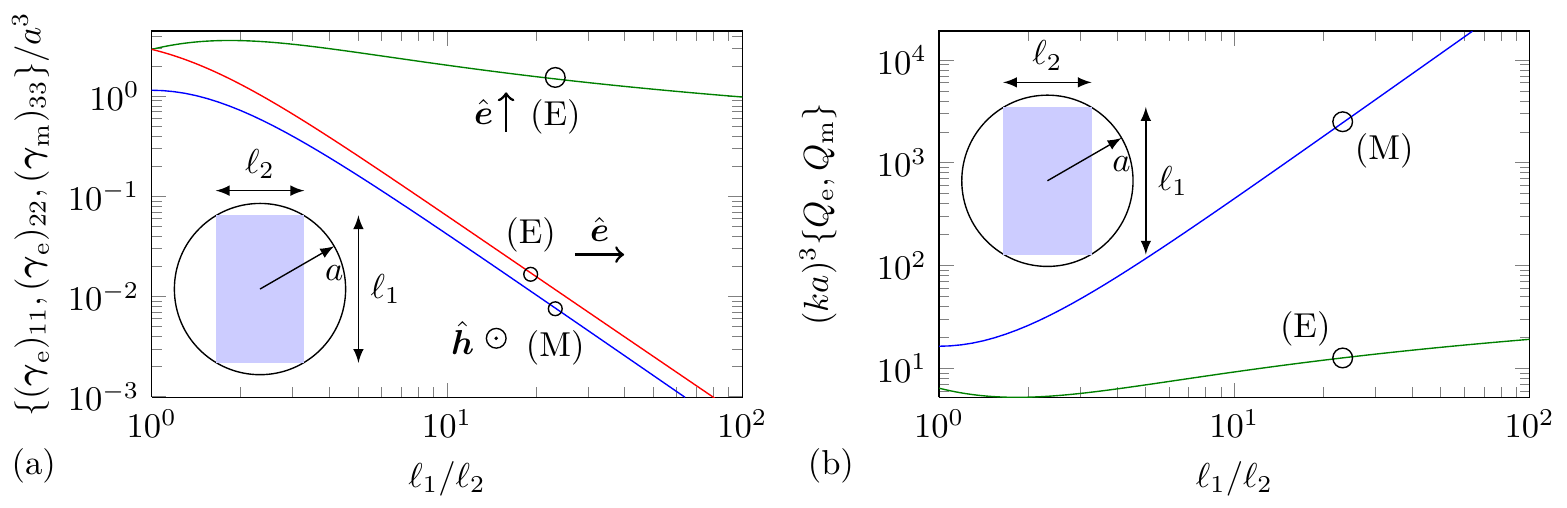}
\caption{(a) Eigenvalues for the electric and magnetic polarizability for an infinitesimally thin plate normalized by $a^3$, where $a=(\sqrt{\ell_1^2+\ell_2^2})/2$ is the radius of a circumscribed sphere. The polarization directions are indicated by $\he$ and $\hh$ for the $\vE$ and $\vH$-fields respectively. The curves are marked with (E) for electric polarizability or (M) for magnetic polarizability. The curves are symmetric with respect to $\ell_1/\ell_2=1$, the lowest curve is the single non-zero eigenvalue of $\gam$. (b) The corresponding $Q$-value from \eqref{Qeev}, \eqref{Qmmax}, once again the (E) correspond to the electric and (M) to the magnetic case.}\label{fig:plate}
\end{figure}

For the sphere we find $(ka)^3\Qe=\frac{3}{2}$ and for a disc $(ka)^3\Qe=\frac{9\pi}{8}$ for the electrical dipole case, see~\ref{app:ell}.
If we instead study $\gae$ of a rectangular plate of size $\ell_2\times \ell_1$ and sweep the ratio $\ell_1/\ell_2$ we find that the two non-zero eigenvalues depicted as the two curves with highest value (red, green) in Figure~\ref{fig:plate}a, and corresponding $Q$ in Figure \ref{fig:plate}b marked with (E). Note that $D/Q=k^3 \he^*\cdot\gae\cdot\he/4\pi$, and hence proportional to the two electric polarizability curves given in Figure~\ref{fig:plate}a, for given direction $\he$. The electrical polarizability here can physically be thought of as how well a structure allow charge separation, in the sense that large eigenvalues in a direction correspond to large static electric dipole-moment, or equivalently large ability to separate charges. 

The corresponding, electric charge maximization problem of $D/Q$ is solved in~\cite{Gustafsson+etal2007a,Gustafsson+etal2009a,Gustafsson+etal2012a}. We have thus the solution to both the $\min_{\rho} Q$ and the $\max_\rho D/Q$ problems for small antennas for small antennas that radiate as electric dipoles. 

\subsection{Antenna Q for an electric current magnetic dipole}\label{sQm}

Analogous to how the electric dipole, $\roe^{(1)}$, and the magnetic dipole with $\rom^{(1)}$ yield the same optimization problem in the previous section, we see that an electric $\vJe^{(0)}$ or a magnetic $\vJm^{(0)}$ current density result in identical optimization problems. We associate a magnetic dipole moment $\vm=m\hatm$ an electric current density, $\vJe^{(0)}$ here denoted $\vJ$, to find the minimization problem:
\begin{equation}\label{amQ}
\Qm=\minimize_{\vJ} \frac{\Wmz(\vJ)}{\Pm(\vJ)} = \minimize_{\vJ} \frac{6\pi}{k^3}\frac{ \int_V \int_V \frac{\vJ^*(\vr_1)\cdot \vJ(\vr_2)}{4\pi|\vr_1-\vr_2|}\diff V_1\diff V_2}{ \big|\int_V \frac{1}{2}\vr\times \vJ\diff V \big|^2},
\end{equation}
with the constraint that $\hn\cdot \vJ=0$ over the surface and $\vJ\in X_0$, as defined in~\eqref{X0} see Section~\ref{sec:emQ} and~\ref{PE} for a more detailed discussion of this choice. This problem is associated with an antenna that radiates as a magnetic dipole \ie $\Pe=0$, and $\We\leq \Wm$. Once the optimization is done we can tune the antenna with a tuning circuit to reach resonance $\We=\Wm$.

We apply once again the method in~\eqref{mQ}--\eqref{eQf} to the minimization of~\eqref{amQ}. Scaling invariance is broken by the assumption that $|\frac{1}{2}\int_V \vr\times\vJ\diff V|=m$, which reduces the problem~\eqref{amQ} to an equivalent problem with Lagrange multipliers. The Lagrangian is
\begin{equation}
\QQ(\vJ,\vJ^*,\lambda_1) =  \int_V \int_V \frac{\vJ^*(\vr_1)\cdot \vJ(\vr_2)}{4\pi|\vr_1-\vr_2|}\diff V_1\diff V_2 -\lambda_1( \big|\int_V \frac{1}{2}\vr\times \vJ\diff V \big|^2 -m^2) 
\end{equation}
for $\vJ\in X_0$, see~\eqref{X0}. The associated critical point equation is 
\begin{align}\label{cj}
\int_V \frac{\vJ(\vr_2)}{4\pi|\vr_1-\vr_2|}\diff V_2 = -\frac{\lambda_1}{2}\vr_1\times  \int_V \frac{1}{2}\vr_2\times\vJ(\vr_2)\diff V_2 = -\frac{\lambda_1 m}{2}\vr_1\times \hatm.
\end{align}
Similar to the electric case~\eqref{lambdaQ} we take the scalar product of \eqref{cj} with $\vJ^*$ and integrate over $V$ to find that $\Qm$ is determined by $\lambda_1$. 
\begin{equation}
\Qm = \frac{6\pi}{k^3}\min_{\vJ} \lambda_1 .
\end{equation}
By applying the operator $\nabla\times\nabla\times$ to~\eqref{cj}, we realize that the currents have support only on the boundary, \ie $\vJ\diff V = \Js\diff S$, and the equation~\eqref{cj} reduce to 
\begin{align}\label{cjs}
\hn \times \int_{\partial V} \frac{\Js(\vr_2)}{4\pi|\vr_1-\vr_2|}\diff S_2 = \frac{\lambda_1 m}{2}\hn\times(\hatm\times \vr_1), \ \text{for}\ \vr_1\in \partial V,
\end{align}
where $\hn$ is normal to $\partial V$. 

Similarly to the electric case~\eqref{EL2}, we compare this with the definition of the magnetic polarizability tensor, $\gam$ in~\eqref{vJ} and \eqref{magpol}: $\gam\cdot \hh H_0 = \vm$. The magnetic polarizability tensor is known, once the region $V$ is given. The $(\lambda_1,\hatm)$ in equation~\eqref{cjs} is hence subject to constraint:
\begin{equation}
\gam\cdot \hatm = \frac{1}{\lambda_1}\hatm.\label{hatm}
\end{equation}
The eigenvalue solution $(\lambda_1,\hatm)$ of~\eqref{hatm} yields the solution to the minimization problem:
\begin{equation}\label{Qmmax}
\Qm = \frac{6\pi}{k^3}\min_{\vJ} \lambda_1 = \frac{6\pi}{k^3(\gam)_3},
\end{equation}
where $(\gam)_3$ is the largest eigenvalue of $\gam$. The analogous case for $D/Q$ is given in~\cite{Gustafsson+etal2012a}. The sphere has magnetic polarizability $2\pi a^3\bm{I}$, which yields $(ka)^3\Qm=3$ \cf~\cite{Thal2006,Pozar2009}. Here $\bm{I}$ is a unit 3 times 3 tensor.

The electric and magnetic polarizabilities of a rectangular plate are depicted in Figure~\ref{fig:plate}a marked with (E) and (M) respectively. The polarizability tensor are diagonal for geometries with two orthogonal reflection symmetries and co-aligned coordinate system~\cite{Payne1956,Kleinman+Senior1986} and for planar structures we have only one eigenvalue of $\gam$, orthogonal to the plane. We can physically think of large $\gam$-eigenvalues as that the region $V$ support a large loop current for the corresponding dipole-moment. Note that planar structures have one non-zero eigenvalue in $\gam$ which is associated to the normal-to-the-surface dipole-moment with the planar `current loop area'. We observe that the magnetic polarizability tensor is connected to the scalar Neumann-problem of Laplace equation, see ~\ref{PE}, and is hence the second of the two `first-moment' (or dipole) quantities associated with a given shape. Note that the magnetic polarizability correspond to the permeable case of $\mu\rightarrow 0$. There are different sign conventions for $\gam$, however we note that $\lambda_1\geq 0$ in ~\eqref{Qmmax} independently of choice of sign-convention in the definition of $\gam$, see~\eqref{amQ}. 

\begin{figure}[!htb]
\begin{center}\includegraphics{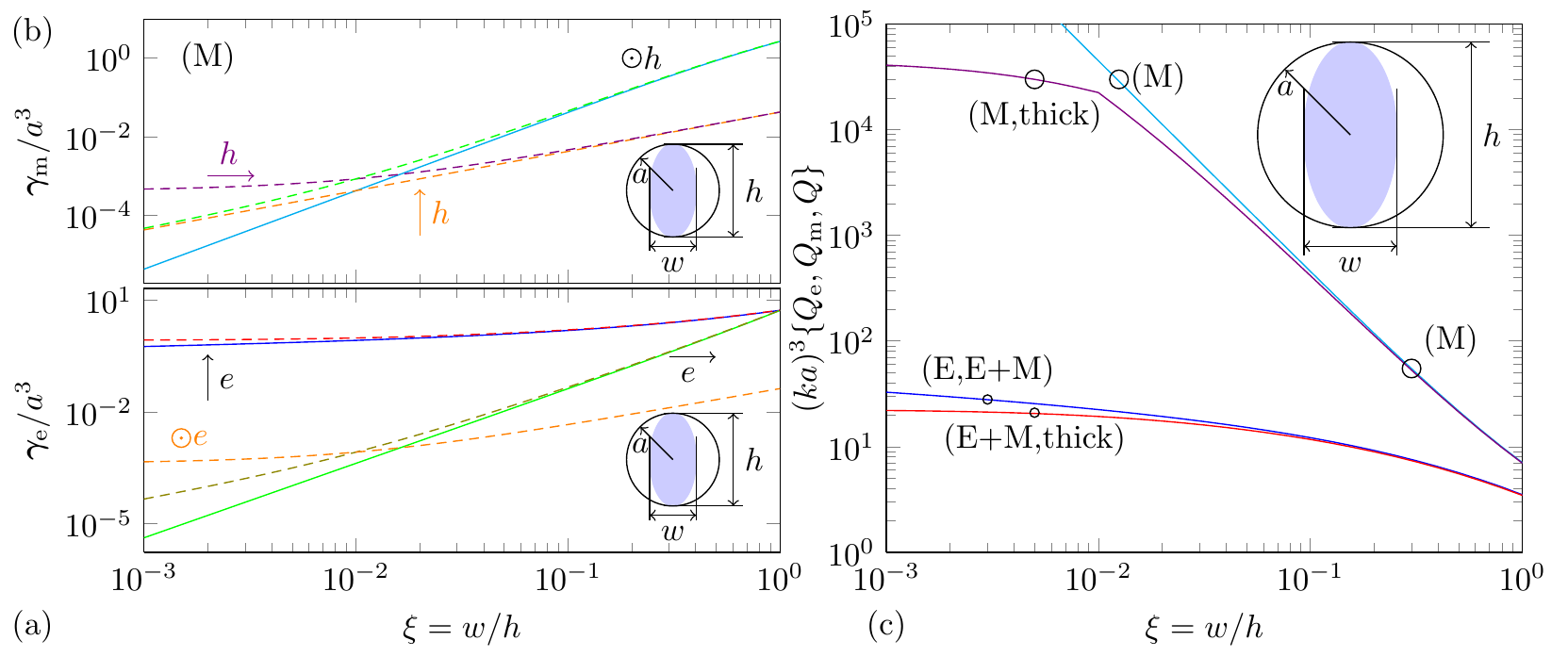}\end{center}
\caption{(a) Eigenvalues to the electric polarizability tensor, $\gae$. Solid lines are the flat-case, dashed lines correspond to the case with normal (out of the paper) radius of the ellipsoidal is $h/100$. Polarization direction is indicated with an arrow. (b) Eigenvalues to the magnetic polarizability tensor $\gam$. Solid line correspond to the flat case, dashed lines are the case with normal radius $h/100$. Note that the $x$-axis is the same as in (a). (c) The antenna $Q$ for a flat ellipse indicated by (E), and (M) and (E+M) corresponding to $\Qe$ from electric sources~\eqref{Qeev}, $\Qm$ from magnetic sources~\eqref{Qmmax} and, $Q$ from combined dual-mode in~\eqref{result} respectively. Two lines are also marked with `thick', to indicate that the ellipsoidal radius normal to the ellipse-surface in the figure is $h/100$. Note in particular for $\Qm$, that as the width becomes smaller than $h/100$, the thickness become important, as is clear in~\eqref{Qmmax}, since it implies a switch of dominant eigenvalue. The reduction of $Q$ as compared to $\Qe$ due to the eigenvalue of $\gam$ is absent for flat structures since the non-zero eigenvalues of $\gae$ and $\gam$ have orthogonal directions. It is a marginal reduction for structures with small thickness. See also~\ref{app:ell}.}\label{fig:flatE}
\end{figure}

A similar current loop-area argument is illustrated in Figure~\ref{fig:flatE}, for a flat ellipse and a thin ellipsoid. The eigenvalues of the polarizability tensor of an ellipsoid are known, see~\ref{app:ell}, and they are depicted in Figure~\ref{fig:flatE}ab. The two curves marked with (M) in Figure~\ref{fig:flatE}c correspond to $\Qm$, the upper one (blue) is for an ellipse of zero thickness and only one $\gam$-eigenvalue corresponding to a current loop-area over the surface. The other marked (M,thick) corresponds to an ellipsoid identical to the flat one, but where the radius normal to the paper is $h/100$ where $h$ is the height of the ellipse. The two transverse eigenvalues of $\gam$ are ignored by $\Qm$ until the width, $w$,  is $h/100$, where equivalent current loop-area of the height-normal (out of the paper) loop dominates the transverse current loop-area and $\Qm$ changes slowly for $w/h<10^{-2}$ since this area is essentially preserved. 

\subsection{Lower bound on antenna Q for both electric charge and magnetic currents}\label{sec:emQ}

The common electric and magnetic dipoles cases above agree with previously derived results~\cite{Gustafsson+Jonsson2013,Yaghjian+etal2013}. We here extend these results to include both the electric charge density $\roe$ and the magnetic current density $\vJm$ \ie the components making up a generalized electric dipole-moment $\de$ \eqref{genpm}. We once again consider the case where the antenna radiates as an electrical dipole, \ie $\Pm=0$ and where the stored energy is mainly electric, $\Wm\leq \We$. After the optimization we tune the antenna to make the stored electric and magnetic energies equal. Optimizing for the $(\Pm,\Wm)$-case is identical to the $(\Pe,\We)$-case up to a sign and the free-space impedance normalization of the currents.  Similar to the above discussion in Section~\ref{sQe} and Section~\ref{sQm} of electric and magnetic dipoles we optimize  
\begin{equation}\label{adQ}
Q=\frac{6\pi}{k^3}
\min_{\rho,\vJ}  \frac{\int_V \int_V \frac{\rho^*(\vr_1)\rho(\vr_2)+\vJ^*(\vr_1)\cdot \vJ(\vr_2)}{4\pi|\vr_1-\vr_2|}\diff V_1\diff V_2}
{\big|\int_V \vr \rho - \frac{1}{2}\vr\times \vJ\diff V \big|^2}.
\end{equation}
We above use the short hand notation $\vJ=\vJm^{(0)}/\eta$, and $\rho=c\roe^{(1)}=\ju \Div\vJe^{(1)}$. 
Here we also have the constraints $\int_V \rho\diff V = 0$ and that $\Div\vJ=0$ to account for the  Gauge-freedom of the associated vector-potential. To include this Gauge-freedom into the optimization problem we restrict the current-density space to $\vJ\in X_0$, see~\eqref{X0}. 

The minimization problem is scaling invariant under transformations $(\rho,\vJ)\mapsto (\rho,\vJ)\alpha$ for any complex valued scalar $\alpha$. By assuming that the denominator has a given value $\lde^2$, we may equivalently consider the problem
\begin{align}
\minimize_{\rho,\vJ} & \int_V \int_V \frac{\rho^*(\vr_1)\rho(\vr_2)+\vJ^*(\vr_1)\cdot \vJ(\vr_2)}{4\pi|\vr_1-\vr_2|}\diff V_1\diff V_2,  
\\ \subto & \big|\int_V \vr \rho(\vr) - \frac{1}{2}\vr\times \vJ(\vr)\diff V \big|^2 = \lde^2, \\ & \int_V \rho^*(\vr) \diff V = 0, \\  & \vJ\in X_0 .
\end{align} 
Using the method of Lagrange multipliers $\lambda_1,\lambda_2$, we define the Lagrangian 
\begin{equation}
\QQ =  \int_V \int_V \frac{\rho^*(\vr_1)\rho(\vr_2)+\vJ^*(\vr_1)\cdot \vJ(\vr_2)}{4\pi|\vr_1-\vr_2|}\diff V_1\diff V_2 -\lambda_1( \big|\int_V \vr \rho - \frac{1}{2}\vr\times \vJ\diff V \big|^2 -\lde^2) - \lambda_2 \int_V \rho^*  \diff V.
\end{equation}
Critical points of $\QQ$ are determined by the variation (Fr\'echet derivative) of $\QQ$. Variation with respect to the Lagrange parameters $\lambda_1$ and $\lambda_2$ gives the constraints. The variation with respect to $\rho^*$ and $\vJ^*$ yields:
\begin{align}\label{el1a}
\int_V \frac{\rho(\vr_2)}{4\pi|\vr_1-\vr_2|}\diff V_2 = \lambda_1\big[\vr_1\cdot \int_V \vr_2 \rho(\vr_2) -\frac{1}{2}\vr_2\times\vJ(\vr_2)\diff V_2\big] + \lambda_2, \\
\int_V \frac{\vJ(\vr_2)}{4\pi|\vr_1-\vr_2|}\diff V_2 = \frac{\lambda_1}{2}\vr_1\times  \int_V \vr_2 \rho(\vr_2) -\frac{1}{2}\vr_2\times\vJ(\vr_2)\diff V_2\label{el1b}.
\end{align}
Here we utilized that $\hn\cdot \vJ=0$ on $\partial V$, and we recognize $\lambda_2$ as a way to ensure that the total charge is zero. To investigate the properties of these Euler-Lagrange equations, we first note that the inner product of these equations with $\rho^*$ and $\vJ^*$ respectively and that their sum can be rewritten as the original problem:
\begin{equation}
Q=\frac{6\pi}{k^3}\min_{\rho,\vJ} \frac{\int_V \int_V \frac{\rho^*(\vr_1)\rho(\vr_2)+\vJ^*(\vr_1)\cdot \vJ(\vr_2)}{4\pi|\vr_1-\vr_2|}\diff V_1\diff V_2}
{\big|\int_V \vr \rho - \frac{1}{2}\vr\times \vJ\diff V \big|^2} = \frac{6\pi}{k^3}\min_{\rho,\vJ} \lambda_1.
\end{equation}
The minimization problem is thus reduced to finding $\lambda_1$ for $\rho,\vJ$ that solves~\eqref{el1a} and \eqref{el1b}. 

Similar to the charge-density case~\eqref{lambdaQ}, we note that $\lambda_1$ implicitly depend on $\rho$ and $\vJ$ through the Euler-Lagrange equations. Another property of the minimization problem appears if we for $\vr\notin \partial V$ operate with $\Laplace$ and with $\Rot\Rot$ on ~\eqref{el1a} and~\eqref{el1b} respectively. We find that $\rho$ and $\vJ$ only have support on the boundary, and we use the notation $\vJ\diff V={\Js}\diff S$ and $\rho\diff V=\rho_s\diff S$. We hence find the Euler-Lagrange equations 
\begin{align}\label{el2a}
\int_{\partial V} \frac{\rho_s(\vr_2)}{4\pi|\vr_1-\vr_2|}\diff S_2 &= \lambda_1(\vr_1\cdot \int_{\partial V} \vr_2 \rho_s(\vr_2) -\frac{1}{2}\vr_2\times{\Js}(\vr_2)\diff S_2) + \lambda_2  = -\lambda_1 \vr_1 \cdot \tde+\lambda_2,\\
\hn_1\times \int_{\partial V} \frac{{\Js}(\vr_2)}{4\pi|\vr_1-\vr_2|}\diff S_2 &= \lambda_1 \hn_1\times (\frac{1}{2}\vr_1\times  \int_{\partial V} \vr_2 \rho_s(\vr_2) -\frac{1}{2}\vr_2\times\Js(\vr_2)\diff S_2)   \nonumber
\\& = -\lambda_1 \hn_1 \times (\frac{1}{2}\vr_1 \times \tde), \label{el2b}
\end{align}
for $\vr_1\in \partial V$. 
We have here introduced the electric and magnetic dipole-moments for the current and charge-distribution that solve~\eqref{el2a} and~\eqref{el2b}:  $\vp=\int_{\partial V} \vr\rho_s\diff S$, $\vm=\frac{1}{2}\int_{\partial V} \vr\times {\Js}\diff S$ and $\tde=\vm-\vp$. 
However both $\vm$ and $\vp$ are presently unknown apart from the constraints that $|\tde|=|\vp-\vm|=\lde$. 

To determine $\lambda_1$, we recall the definitions of the electric polarizability tensor $\gae$ and magnetic polarizability tensor $\gam$ in~\ref{PE}. We compare~\eqref{el2a} and~\eqref{el2b} with \eqref{pe} and \eqref{vJ}. The polarizability tensors $\gae$ and $\gam$ are known, once $V$ is given, and we find that~\eqref{Dgae} and~\eqref{magpol} impose constraints on $\lambda_1$ and $\tde$:
\begin{equation}
\gae \cdot(\vp-\vm) = \frac{1}{\lambda_1} \vp, \ \
\gam \cdot (\vp-\vm) = \frac{-1}{\lambda_1} \vm.
\end{equation}
Adding the two equations yields that $\lambda_1^{-1}$ is an eigenvalue to the matrix $\gae + \gam$. Furthermore, $\vm-\vp = \lde\hde$, where $\hde$ is an eigenvector of $\gae+\gam$ of unit length. Thus we have found that in this case the lower bound on $Q$ is given by
\begin{equation}\label{result}
Q = \frac{6\pi}{k^3(\gae+\gam)_3},
\end{equation}
where $(\gae+\gam)_3$ is the largest eigenvalue of the $\gae+\gam$ tensor. The corresponding $\rho_s,{\Js}$ are hence the solution of~\eqref{el2a} and \eqref{el2b}, where $\vm-\vp=\lde\hde$, \ie in the direction of the unit eigenvector corresponding to the largest eigenvalue. This result is similar to~\cite{Yaghjian+etal2013}, but derived with a different method. Note that $\gae+\gam\geq 0$.

The minimization procedure also establish that there exists a $\lambda_1\geq 0$ such that
\begin{equation}\label{ineq}
\big| \int_V \vr \rho - \frac{1}{2}\vr\times \vJ \diff V\big|^2 \leq \frac{1}{\lambda_1} \int_V\int_V \frac{\rho^*(\vr_1)\rho(\vr_2) + \vJ^*(\vr_1)\cdot\vJ(\vr_2)}{4\pi|\vr_1-\vr_2|}\diff V_1\diff V_2 
\end{equation}
for all $\rho$ and $\vJ$ that satisfy the bi-condition $\vJ\in X_{0}$  and $\int_V \rho\diff V = 0$. Equality is reached when $\rho$ and $\vJ$ satisfy the Euler-Lagrange equations above, yielding $1/\lambda_1 = (\gae+\gam)_3$. 
An equivalent formulation of this result is 
\begin{equation}
\Pe \leq (\gae+\gam)_3 \frac{c k^4}{3\pi} \We,\ \text{or}\ 
\Pm \leq (\gae+\gam)_3 \frac{c k^4}{3\pi} \Wm,  
\end{equation}
for the above described currents. The identity is achieved in either case for currents that realize the minimization of $\Qe$ or $\Qm$. The inequality for the $(\Pm,\Wm)$-case is obtained identically with above described case starting from $\Pm$ and $\Wm$ with the substitution of $\vJ=-\vJe^{(1)}$ and $\rho=c\rom^{(1)}/\eta$ giving \eqref{adQ} with $\vr\rho + \vr\times \vJ/2$ of the integrand in the denominator.

\subsubsection{Comparisons and numerical examples for the $Q$-lower bound for the dual-mode case~\eqref{result}} 
\begin{figure}[!htbp]
\begin{center}
\includegraphics{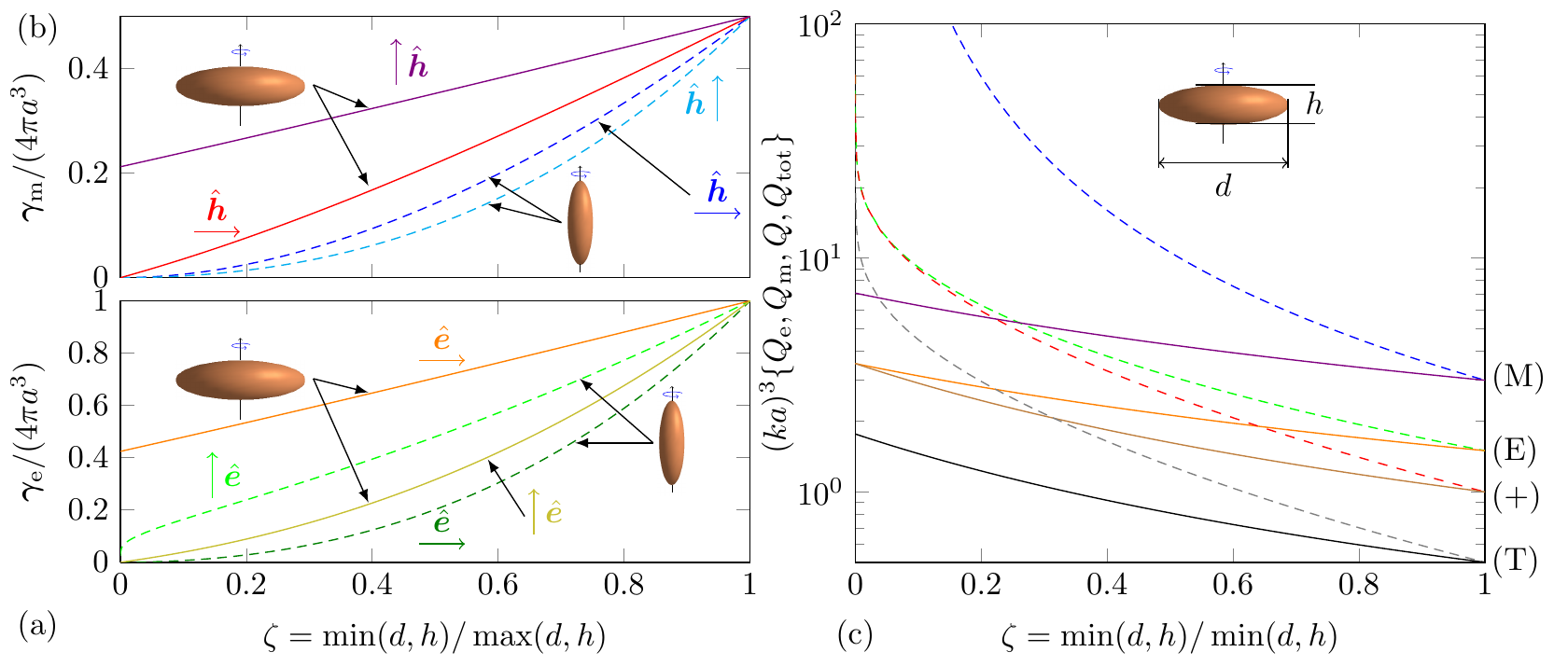}
\end{center}
\caption{(a) Eigenvalues for the electric polarizability tensor. Polarization direction is indicated with a vertical or horizontal arrow. (b) Eigenvalues of the magnetic polarizability tensor. The $x$-axis is the same as the axis in (a). (c) $\Qe$ from \eqref{Qeev} is marked with (E), $\Qm$ from~\eqref{Qmmax} marked with (M) and $Q$ from~\eqref{result} are marked with a (+) and for the dual-mode antenna~\eqref{gen} with a (T). Note that the $(\gae)_{11}=2(\gam)_{33}$, as for axial-symmetric objects shown in~\cite{Payne1956}. Dashed lines in (abc) are the prolate case, solid lines are the oblate case. $(ka)^3(\Qm,\Qe,Q_{+},Q_{T})\rightarrow (3,3/2,1,1/2)$ as $\zeta\rightarrow 1$ \ie the sphere. See \ref{app:ell}.}\label{fig:gam}
\end{figure}

We note that for a sphere where both electric and magnetic currents contribute to the generalized electric dipole-moment we find that $(ka)^3\Qe=1$~\cite{Chu1948}. The $Q$-lower bound for the flat ellipse and the thin ellipsoid are depicted in Figure~\ref{fig:flatE}. 
For planar structures we note that there is only one non-zero eigenvalue of $\gam$, in the direction normal to the surface and hence perpendicular to the non-zero direction of $\gae$. For a rectangular plate this eigenvalue is depicted in Figure~\ref{fig:plate}. We conclude that in planar structures $\gae$ and $\gam$ do not couple to improve the antenna $Q$. As is clear from the case where we add a small thickness of the domain as in Figure~\ref{fig:flatE}c, we see that there is a rather small reduction of $Q$ as compared with the flat case.

The polarizability tensors for spheroidal shapes are known, see~\ref{app:ell} and Figure~\ref{fig:gam}ab. We depict $Q$ for spheroidal bodies as a function of the ratio between height and diameter in Figure~\ref{fig:gam}c. Here, the curve marked with (+) correspond to $Q$ given in~\eqref{result} are shown for both the prolate (dashed lines) and oblate cases (solid lines). 

\begin{figure}[!htb]
\begin{center}
\includegraphics{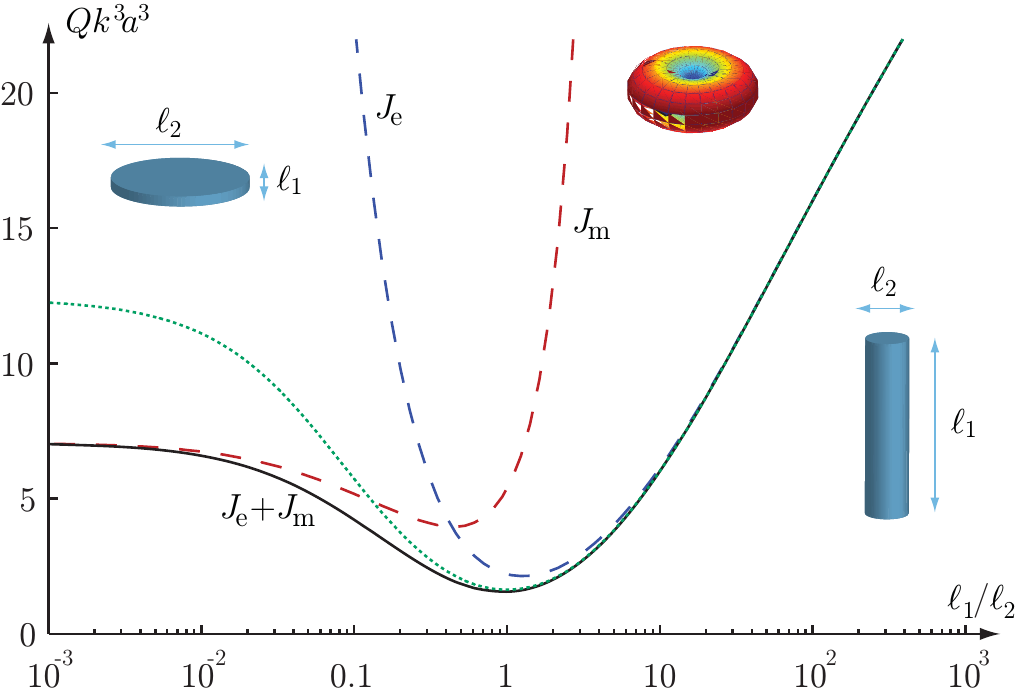}
\end{center}
\caption{The figure depicts the antenna Q, for energies that corresponds to currents that radiate as an electrical dipole aligned with the vertical axis. The dotted green line correspond to $Q_V$ in~\eqref{QV}, the $\vJe$, $\vJm$ and $\vJe+\vJm$ correspond to $\Qe$, $\Qm$ and $Q$ in respectively \eqref{Qeev}, \eqref{Qmmax} and \eqref{result}.}\label{cyl} 
\end{figure}

The approach in~\cite{Yaghjian+Stuart2010} provides an  antenna Q, $Q_V$, depending only on $\gae$ and volume $V$. To compare $Q_V$ with~\eqref{result} we use the inequality~\cite[1.5.19]{Schiffer+etal1949}:
\begin{equation}
(\he\cdot\gae\cdot\he-V)(\he\cdot\gam\cdot \he-V) \geq V^2 \Leftrightarrow 
(\he\cdot\gae\cdot\he-V)(\he\cdot(\gae +\gam)\cdot \he) \geq (\he\cdot\gae\cdot\he)^2 .
\end{equation}
Rewriting and comparing with the results we find that 
\begin{equation}\label{QV}
Q=\frac{6\pi}{k^3 \he\cdot (\gae+\gam)\cdot\he} \leq \frac{6\pi}{k^3\he\cdot\gae\cdot\he}(1-\frac{V}{\he\cdot\gae\cdot\he}) = Q_V,
\end{equation}
if we choose the $\he$ to be the unit eigenvectors corresponding to the largest eigenvalue of $\gae+\gam$. Equality holds for several cases in particular for ellipsoidal-shapes. An updated approach to antenna Q is given in~\cite{Yaghjian+etal2013}, see also~\cite{Jonsson+Gustafsson2013a}. 
To illustrate that there is a difference between $Q$ and $Q_V$ we calculate both antenna Q's for a cylinder. We assume here that the currents radiate as an electrical dipole aligned with the cylinder axis, \ie the vertical $\hx_3$-axis, the resulting $Q$ from ~\eqref{result} and $Q_V$ are shown in Figure~\ref{cyl}. 
To demand that a small antenna radiates as an electric dipole in a given direction is equivalent with selecting the corresponding eigenvalue of the polarizability tensor. Such a choice of eigenvalue does not necessarily minimize antenna $Q$.

The above examples illustrate how the shape of a small antenna enters into the antenna Q-bound. The shape characterization in antenna Q is encoded in the respective polarizability tensors. The electric polarizability is a measure on how easy it is to separate charge for a given $V$, \ie to create a large electric dipole-moment. Similarly, the magnetic polarizability measure how easy it is to create a large magnetic dipole moment, \ie finding a large `current-loop area' in the domain.

If we similarly to~\cite{Hansen+etal2012,Stuart+Yaghjian2010,Yaghjian+etal2013} associate the magnetic currents with layers/volumes of magnetization or synthesized Amperian current loops we note that the associated volumes for the electric and magnetic currents do not necessary need to occupy identical volumes/surfaces. In such a case there are a considerable design freedom for $\gae$ and $\gam$, with the performance bounded by the eigenvalues of $\gae+\gam$ for the total volume $V$. 

\subsection{Dual mode antennas}\label{dual}

Self-resonant dual mode-antennas where both the electric $\Pe$ and magnetic $\Pm$ dipole radiation contribute significantly to the radiation and $\We=\Wm$ is considered here.  Utilizing that the problem decouples, we use the respective electric and magnetic case above with identities \eqref{ineq} where $\lambda_1\geq 0$  for both $\We$ and $\Wm$. We hence find that the general case can be bounded by:
\begin{equation}\label{gen}
Q \geq \frac{6\pi}{k^3}\frac{\max(\We,\Wm)}{\lambda_1^{-1} \We + \lambda_1^{-1}\Wm} \geq \frac{6\pi}{k^3}\frac{\lambda_1}{2} = \frac{3\pi}{k^3(\gae+\gam)_3},
\end{equation}
which follows directly from the H\"older inequality~\cite{Lieb+Loss1997}. Equality follows when both electric and magnetic charges are optimized and the antenna is self-resonant. Clearly we find that $Q$ is half the value of $\Qe$ or $\Qm$ when only electric or magnetic dipole radiation is allowed. The sphere yields $(ka)^3Q=1/2$, which agrees with the result of the sphere given in~\cite{Chu1948,Harrington1960,Mclean1996}. A similar result is given in~\cite{Yaghjian+etal2013}, derived with a different method\footnote{
Optimization that utilize a fixed electric to magnetic dipole radiation ratio is discussed in~\cite{Yaghjian+etal2013}.}. The antenna $Q$ for this case is illustrated for spheroidal shapes in Figure~\ref{fig:gam}c, for curves marked with a (T).

\section{Convex optimization for optimal currents}\label{4}
Bounds on $D/Q$ can be expressed as a convex optimization problems~\cite{Gustafsson+Nordebo2013}. Here, these results are generalized to include electric and magnetic current densities.  
We consider a volume $V$ with electric $\vJe$ and magnetic $\vJm$ current densities. We expand the current densities in local basis-functions
\begin{equation}\label{eq:basis}
	\vJe(\vr)\approx \sum_{n=1}^N J_{\mrm{e},n}\psiv_n(\vr)
	\text{ and }
	\vJm(\vr)\approx \sum_{n=1}^N J_{\mrm{m},n}\psiv_n(\vr)
\end{equation}
and introduce the $1\times 2N$ matrix $\vJv$ with elements $\{J_{\mrm{e},n}\}$ for $n=1,...,N$ and $\{\eta^{-1}J_{\mrm{m},n-N}\}$ for $n=N+1,...,2N$ to simplify the notation. The basis functions are assumed to be real valued, divergence conforming, and having vanishing normal components at the boundary~\cite{Peterson+Ray+Mittra1998}. 

A standard method of moment implementation using the Galerkin procedure computes the stored energies given in~\ref{app:result} as matrices. For simplicity, we here compute these stored energy matrices $\Xme$ and $\Xmm$ only for the leading order term in \eqref{Le} and \eqref{Lm}, for $ka\ll 1$, \ie
\begin{equation}
	X^{\mrm{e}}_{ij} 
	=\frac{1}{k}\int_{V}\!\!\int_{V}\nabla_1\cdot\psiv_{i}(\vr_1)\nabla_2\cdot\psiv_{j}(\vr_2)
\frac{\cos(kR_{12})}{4\pi R_{12}} \diff V_1\diff V_2
\label{eq:EFIEe}
\end{equation} 
and
\begin{equation}
	X^{\mrm{m}}_{ij} 
	=k\int_{V}\!\!\int_{V}\psiv_{i}(\vr_1)\cdot\psiv_{j}(\vr_2)
\frac{\cos(kR_{12})}{4\pi R_{12}} \diff V_1\diff V_2.
\label{eq:EFIEm}
\end{equation} 
The quadratic forms for the stored energies \eqref{we} and \eqref{wm} are then approximated as
\begin{equation}
	\We\approx\frac{\eta}{4\omega}\vJv^{\herm}\Xme\vJv
	=\frac{\eta}{4\omega}\sum_{i,j=1}^N J_{\mrm{e},i}^\ast X^{\mrm{e}}_{ij}J_{\mrm{e},j} + J_{\mrm{m},i}^\ast X^{\mrm{m}}_{ij}J_{\mrm{m},j}
\label{eq:aa}
\end{equation}
and
\begin{equation}
	\Wm\approx\frac{\eta}{4\omega}\vJv^{\herm}\Xmm\vJv
	=\frac{\eta}{4\omega}\sum_{i,j=1}^N J_{\mrm{e},i}^\ast X^{\mrm{m}}_{ij}J_{\mrm{e},j} + J_{\mrm{m},i}^\ast X^{\mrm{e}}_{ij}J_{\mrm{m},j}.
\label{eq:ab}
\end{equation}
where the superscript, $\herm$, denotes the Hermitian transpose.  

We also use the radiated far field, $\Eo(\hr)$ see~\eqref{eoo}.
The radiation vector projected on $\he$, \cf~\eqref{eoo}, defines the $2N\times1$ matrix $\Fm$ from
\begin{equation}\label{eq:radvectorMatrix}
	\he^{\ast}\cdot\Eo(\hk)
	\approx
	\Fm\vJv
	=-\ju \eta k\sum_{n=1}^{N} \big[ J_{\mrm{e},n}\int_V\he^{\ast}\cdot\psiv_n(\vr)\frac{\lexp{\ju k\hk\cdot\vr}}{4\pi}\diff V
	+J_{\mrm{m},n}\int_V\hk\times\he^{\ast}\cdot\psiv_n(\vr)\frac{\lexp{\ju k\hk\cdot\vr}}{4\pi}\diff V\big] ,
\end{equation}

Using the scaling invariance of $D/Q$, we rewrite the maximization of $D/Q$ into the convex optimization problem of maximization of the far-field in one direction subject to a bounded stored energy~\cite{Gustafsson+Nordebo2013}, \ie
\begin{equation}\label{eq:GoQ2}
	\begin{aligned}
		& \maximize_{\vJv} && \RE\{\Fm\vJv\}, \\
		& \subto && \vJv^{\herm}\Xme\vJv \leq 1,\\
		&  && \vJv^{\herm}\Xmm\vJv \leq 1.
	\end{aligned}
\end{equation}
The formulation is easily generalized by adding additional convex constraints~\cite{Gustafsson+Nordebo2013}. There are several efficient implementations that solve convex optimization problems, here we use \texttt{CVX}~\cite{Grant+Boyd2011}.

\begin{figure}[!ht]
\begin{center}\includegraphics{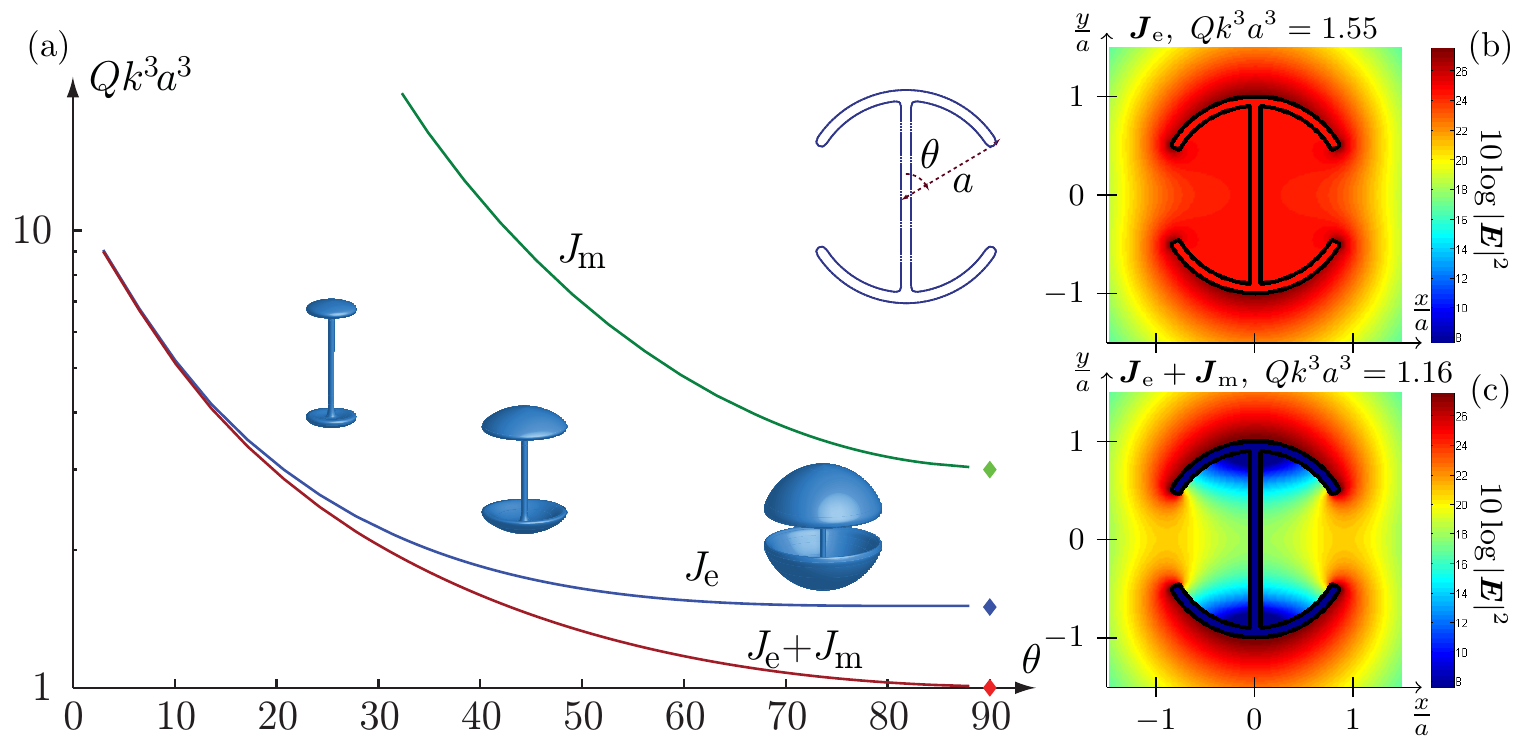}
\caption{(a) The capped spherical dipole. The figure shows the optimized antenna Q for different values of the cap-angle, see the figure in at top right. The purely electric and the purely magnetic cases are shown in blue and green colors. The joint case is given in the red-curve. Note that the constraint of only electrical energy approaches: $\vJe$ yield $\Qe(ka)^{3}=3/2$, $\vJm$ yield $\Qe(ka)^3=3$ and the combined electric case $\vJe$ and $\vJm$ yield $\Qe(ka)^3=1$ as $\theta=90^\circ$. (bc) The figure shows a comparison of the interior field without (b) and with (c) magnetic currents for dipoles that radiate as an electric dipole.
}\label{Capped}
\end{center}
\end{figure}

We consider planar geometries and bodies of revolution to illustrate the bound. The resulting $Q$ of~\eqref{qeqm} for a small spherical capped dipole antenna is depicted in Figure~\ref{Capped}a as a function of the angle $\theta$ for a maximized omnidirectional partial directivity in $\theta=90^\circ$ and polarized in the $\hz$-direction. The resulting radiation pattern is as from a $\hz$-directed electric Hertzian dipole, \ie $D=1.5\sin^2\theta$. The three cases; electric and magnetic currents $\Jve+\Jvm$, only electric currents $\Jve$, and only magnetic currents $\Jvm$ are analyzed. The requirement of electric dipole-radiation implies $\Pm=0$, $\Pe\neq 0$, and that we can use $\roe$ to represent the electric currents $\vJe$. We observe that the $\theta=90^\circ$ case corresponds to a spherical shell with the classical~\cite{Chu1948,Thal2006,Yaghjian+Stuart2010,Stuart+Yaghjian2010} bounds $Qk^3a^3=\{1,1.5,3\}$ for the $\Jve+\Jvm$, $\Jve$, and $\Jvm$ cases, respectively. The reduced $Q$ of the combined $\Jve+\Jvm$ case is understood from the suppression of the energy in the interior of the structure. This is also shown in Figure~\ref{Capped}bc, where the resulting electric energy density is depicted for the cases to electric currents $\Jve$ and combined electric and magnetic currents $\Jve+\Jvm$. We also note that the potential improvement with combined electric and magnetic currents $\Jve+\Jvm$ decreases as $\theta$ deceases. This can be understood from the increased internal energy as the magnetic current can only cancel the internal field for closed structures. Moreover, the faster increase of $Qk^3a^3$ as $\theta\to 0$ for the $\Jvm$ case than for the $\Jve$ case is understood from the loop type currents of $\Jvm$ whereas $\Jve$ is due to charge separation.

\begin{figure}
\begin{center}
\includegraphics{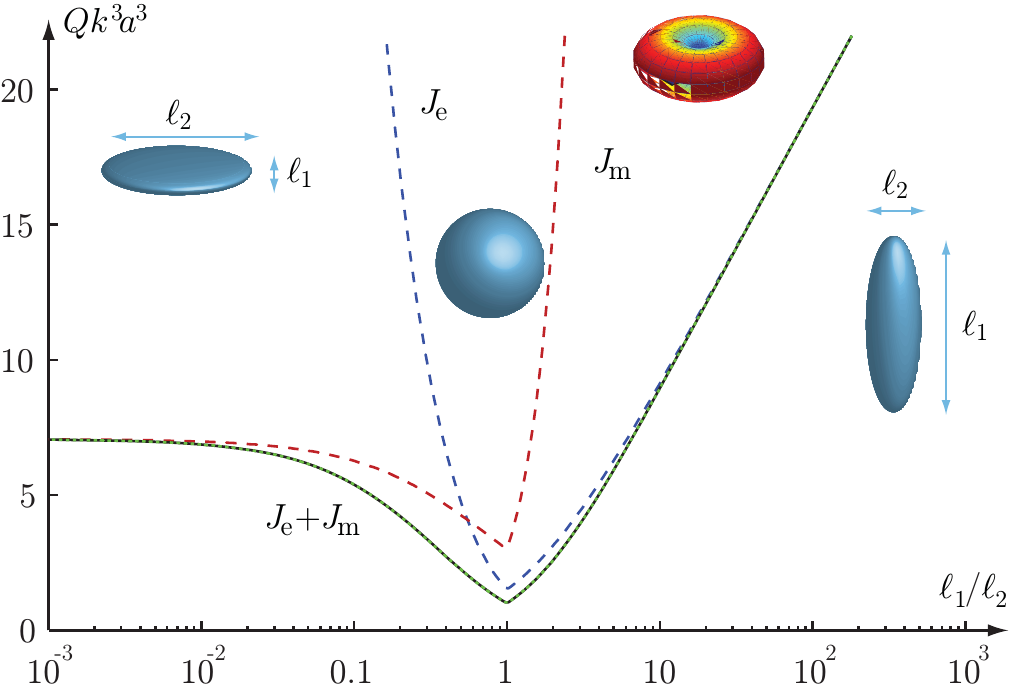}
\end{center}
\caption{Sweeping the two diameters of a spheroid, with purely electric and purely magnetic currents, as well as the combination are shown. Here the optimization is done under the assumption that the far-field radiates as a electric dipole aligned with the vertical axis. See also the discussion at the end of Section~\ref{4}.}\label{fig:spheroid}
\end{figure}

The case of a spheroidal body with the additional radiation constraint corresponding to an electrical dipole along the vertical axis is given in Figure~\ref{fig:spheroid}. It is interesting to compare this constrained result with, with the minimal $Q$ as shown in Figure~\ref{fig:gam}c, the (+)-curve. Small $\ell_1/\ell_2$ in Figure~\ref{fig:spheroid} corresponds to small $\zeta$ with solid lines of Figure~\ref{fig:gam}c. We see that in the constrained case $Q$ approaches the pure magnetic current-case marked $\vJm$, whereas in Figure~\ref{fig:gam}c, $Q$ marked with (+), approaches the pure $\Qe$ case (solid line marked (E)), and it is a lower value than the result indicated in Figure~\ref{fig:spheroid}. The cause of this difference is the requirement of the radiation pattern, locking $Q$ to a disadvantageous eigenvalue, see Figure~\ref{fig:gam}a and the vertical polarization direction (solid line). The physical interpretation is clear: for the disc it is easier to excite an electrical dipoles aligned with the surface. The required vertical electric dipole is the cause of the higher $Q$ in Figure~\ref{fig:spheroid}. For $\ell_1/\ell_2$ large, we see that both results agree (dashed lines in Figure~\ref{fig:gam}c, as $\zeta\rightarrow 0$).

\section{Conclusion}

The present paper introduces a common mathematical framework for deriving lower bounds on antenna Q to arbitrary shapes for electric and magnetic current densities. For the corresponding cases considered in \cite{Gustafsson+etal2007a,Yaghjian+etal2013} we get identical results for appropriate choices of the ratio of electric and magnetic dipole radiation $\Pe$ and $\Pm$. This is rather remarkable since the underlying physics and mathematical approaches utilize widely different ways to arrive to antenna Q and $D/Q$.  The result also verify that both electric and magnetic current densities are required to reach the classical results for a sphere in \eg~\cite{Chu1948,Harrington1960}.  The present method also provides a minimization method to determine the minimizing currents, which is attractive for optimization procedures, where antenna-Q related problems can be considered. A few of these are demonstrated in the present paper, and extensions analogously to the convex optimization results in~\cite{Gustafsson+Nordebo2013} follows directly from the explicit results shown here. 

In the paper we derive the antenna Q lower bound for small electric antennas. The lower bound on antenna Q depends symmetrically on both the electric and magnetic polarizabilities, which reflect the dual symmetry of the electromagnetic equations with electrical and magnetic current densities. The explicit lower bound enables a priori estimates of antenna Q given the shape of the object in terms of the static polarizability tensors $\gae$ and $\gam$. We also determine the antenna Q for planar rectangles, ellipsoids and cylinders. Here we sweep a geometrical shape parameter, to illustrate how the antenna properties $Q$ and $D/Q$ depend on the shape. Low antenna $Q$ is associated with low fields inside closed domains, with the present technique we can study objects like the spherical cap to observe how the cancellation of the fields in the interior of an essentially open structures behave for optimal or constrained antenna $Q$.  

We conclude, that the presented current density representation of the stored energy yields explicit analytical expressions on antenna Q and $D/Q$ in terms of the polarizability tensors. We also illustrate that the polarizabilities and different antenna Q-related optimization problems are straight forward to calculate, given standard software. This follows through the relation of the polarizabilities to the scalar Dirichlet and Neumann problems. The present results are applicable to a range of practical antenna problem, as a priori limitations of their antenna Q-performance, and more subtle as explicit current minimizers that might give insight into antenna design problems.

\appendix

\section{Stored energy -- general sources}\label{app:result}

The stored energies are derived from~\eqref{WEM} using an approach with potentials. The result consist of a sum of terms each of a given leading $k^n$-behavior for $n=0,1,\ldots$ as $k\rightarrow 0$
\begin{equation}\label{A1}
\We= \We^0 + \Wem^1 + \Wem^2+\Wem^3 + \Wem^{\text{rest}}, \ \ 
\Wm = \Wm^0 + \Wem^1+\Wem^2-\Wem^3+\Wem^{\text{rest}}.
\end{equation}
The EFIE operators $\Le$ and $\Lm$ are given in~\eqref{Le} and \eqref{Lm}, and we find the leading order electric and magnetic stored energy as
\begin{equation}
\We^0 = \frac{\mu}{4k}\IM \big[\dotp{\vJe}{\Le\vJe}+\frac{1}{\eta^2}\dotp{\vJm}{\Lm\vJm} \big] \sim \Oh(1), \ k\rightarrow 0
\end{equation}
and
\begin{equation}
\Wm^0 = \frac{\mu}{4k}\IM \big[\dotp{\vJe}{\Lm\vJe}+\frac{1}{\eta^2}\dotp{\vJm}{\Lm\vJm} \big] \sim \Oh(1) , \ k\rightarrow 0.
\end{equation}
Both terms are to leading order 1 for small $k$, as is indicated by the $\Oh(1)$ above.

The second term contains the leading order cross-term:
\begin{equation}
\Wem^1 = \frac{-\mu}{4k\eta}\IM\dotp{\vJe}{\KK_2\vJm}\sim \Oh(k^1) ,
\end{equation}
where
\begin{equation}
\dotp{\vJe}{\KK_2\vJm}=\frac{k^2}{4\pi}\int_V\int_V \vJe^*(\vr_1)\cdot \hR\times \vJm(\vr_2)\cos(k|\vr_1-\vr_2|)\diff V_1\diff V_2.
\end{equation}
Here $\vR=\vr_1-\vr_2$, $R=|\vR|$, and $\hR=\vR/R$. The next higher order term is
\begin{equation}
\Wem^2=\frac{\mu}{4 k}\IM \big[\dotp{\vJe}{\Lem\vJe} + \frac{1}{\eta^2}\dotp{\vJm}{\Lem\vJm}\big] \sim \Oh(k^2),
\end{equation}
where
\begin{equation}
\dotp{\vJ}{\Lem\vJ}=
\ju \int_V\int_V  [k^2 \vJ_{1}\cdot\vJ_{2}^* - (\Div\vJ_{1})(\Div\vJ_{2}^*)] \frac{\sin(k|\vr_1-\vr_2|)}{8\pi} \diff V_1\diff V_2 .
\end{equation}
The $\Wem^3$ term is
\begin{equation}
\Wem^3 = \frac{-\mu}{4\eta k}\RE \dotp{\vJe}{\KK_1\vJm} \sim \Oh(k^3),
\end{equation}
where 
\begin{equation}\label{K1}
\dotp{\vJe}{\KK_1\vJm}= \frac{k^2}{4\pi}\int_V\int_V \vJe^*(\vr_1)\cdot \hR\times \vJm(\vr_2)\sbj_1(kR)\diff V_1\diff V_2.
\end{equation}

The last term $\Wem^{\text{rest}}$ is $\Oh(k^3)$ for small $k$ and it is coordinate dependent in certain cases~\cite{Gustafsson+Jonsson2014} 
\begin{equation}
\Wem^{\text{rest}}= \frac{\mu}{4}\big[K_3(\vJe) + \frac{1}{\eta^2}K_3(\vJm)+ K_4(\vJe,\vJm)\big],
\end{equation}
where
\begin{equation}
K_3(\vJ)= -\int_V \int_V  \IM [k^2 \vJ_{e,1}\cdot\vJ_{e,2}^*  -(\Div\vJ_{e,1})(\Div\vJ_{e,2}^*)] \frac{(r_1^2-r_2^2)}{8\pi R}\sbj_1(kR)   \diff V_1\diff V_2 \sim \Oh(k^4)
\end{equation}
and
\begin{equation}
K_4(\vJe,\vJm)= \frac{k}{\imp} \int_V \int_V
 \RE [\vJ_{m,2}^*\times \vJ_{e,1}] \cdot \\
\big[ \frac{ \vr_2+\vr_1}{4\pi R}\sbj_1(kR) + k\hR\frac{r_1^2-r_2^2}{4\pi R}\sbj_2(kR) \big]\diff V_1\diff V_2 \sim \Oh(k^3).
\end{equation}
Note that both $\Wem^{\text{rest}}$ and $\Wem^3$ are of the same asymptotic order in $k$. We keep the terms separate due to the sign-change of $\Wem^{3}$ in~\eqref{A1} and since $\Wem^{\text{rest}}$ can depend on the coordinate system. We consider the coordinate independent part of these energies as the essential physical quantity of the stored energy.

\section{Polarizability tensors}\label{PE}

The electric and magnetic polarizability tensors are well known in electromagnetic scattering~\cite{Rayleigh1871,Kleinman+Senior1986,deMeulenaere+vanBladel1977,Arvas+Harrington1983}, and since they enter as an essential part in this work, we review their definition and a few different approaches to compute them. The polarizability tensors required in this paper are properties of the geometrical shape $V$ only~\cite{Schiffer+etal1949,Payne1956,Kleinman+Senior1986}, similar to the capacitance. The magnetic polarizability appear also in fluid-dynamics as the virtual mass~\cite{Schiffer+etal1949}. 
We denote the electric and magnetic polarizability tensors with $\gae$ and $\gam$. For sufficiently regular domains it is known~\cite{Schiffer+etal1949} that $\gae$ is associated with a scalar Dirichlet-problem and $\gam$ with a scalar Neumann-problem for the Laplace operator. 

If the shape $V$ has two orthogonal reflection symmetries this reduces $\gae$ and $\gam$ to diagonal matrices for a co-aligned coordinate system~\cite{Kleinman+Senior1986}. If $V$ is axial-symmetric along \eg the $x_3=z$-axis this reduce the number of unknowns to three~\cite{Payne1956}: $(\gae)_{11}=(\gae)_{22}=2(\gam)_{33}$, $(\gam)_{11}=(\gam)_{22}$ and $(\gae)_{33}$, where the index denote the respective matrix element.

\subsection{Electric polarizability tensor}

Here we assume that the boundary $\partial V$ is smooth with a well-defined concept of inside and outside in order to define an outward normal $\hn$, see Figure~\ref{fig1}, and will upon occasion also consider degenerate surfaces like the rectangular plate. For generalizations of the associated potential theory to Lyapunov and Lipschitz surfaces~\cite{Costabel1988,Hsiao+Wendland2008,Helsing+Perfekt2013}.  Consider a perfectly conductive object, $V$ (or a high contrast object) in a homogeneous external electric field $E_0\he$. The external Dirichlet problem for the electric potential has the solution, $\phi_0$ that is related to the perturbed potential $\phi$ through $\phi_0=\phi-E_0\he\cdot \vr$, and we have that 
\begin{align}
\Laplace \phi&=0, \  \vr\in \RR^3\backslash V,\nonumber\\
\phi &= E_0\he\cdot\vr + K, \  \vr\in\partial V, \label{D}\\
\phi &= \Oh(r^{-1}),  \ \text{as}\ |\vr|\rightarrow \infty . \nonumber
\end{align}
The constant $K$ is selected in such a way that the total induced charge $q$ is zero. Here we have $q=\int_{\partial V}\rho_s\diff S$ and $\rho_s=-\eps\partial_n\phi_0=-\eps\partial_n( \phi - E_0 \he\cdot\vr)$ on $\partial V$. The dielectric constant in vacuum is here denoted $\eps$. The system~\eqref{D} is a well-posed problem and there exists a range of algorithms to solve it, like Fredholm integrals of first and second kind~\cite{Folland1995}. 

The polarizability tensor, $\gae$, is a linear map between the boundary condition $\eps E_0 \he$ and the dipole-moment, $\vp=\int\vr \rs\diff S$, see~\cite{Schiffer+etal1949,Kleinman+Senior1986,Gustafsson+etal2012a}. To explicitly find this relation, we connect the potential to the boundary condition through the single layer potential:
\begin{equation}\label{pe}
\eps\phi(\vr_1) = \int_{\partial V} \frac{\rho_s(\vr_2)}{4\pi|\vr_1-\vr_2|}\diff S_2 = D_0\he\cdot \vr_1 + \eps K,\ \vr_1 \in \partial V
\end{equation}
for a given electric displacement field $\vD=D_0\he=\eps E_0 \he=\eps\vE$. It is known~\cite{Kellogg1929} that~\eqref{pe} can be generalized to shapes $V$ like the 2D-plate and other objects with corners. 

A Method of Moments approach can solve this first order integral equation for $\rho_s$, but care is required to account for possible charge-density singularities near corners or edges as well as large condition numbers. The multipole expansion of the potential as $\vr_1\rightarrow \infty$ is given by
\begin{equation}\label{pot}
4\pi\eps\phi(\vr_1) = \int_{\partial V} \frac{\rs(\vr_2)}{|\vr_1-\vr_2|}\diff S_2 \rightarrow \frac{q}{r_1} + \frac{\vp\cdot \hr_1}{r_1^2}+\Oh(r_1^{-3}),\ \text{as}\ r_1\rightarrow \infty.
\end{equation}
The electric polarizability tensor, a $3\times 3$-matrix, $\gae$, is defined as the map
\begin{equation}\label{Dgae}
\gae\cdot \he D_0 = \vp.
\end{equation}
From this definition it follows that the dimension of $\gae$ is volume. A procedure to calculate $\gae$ is to subsequently insert three orthonormal directions as $\he$ in the boundary condition in~\eqref{D}, and for each of these, find the corresponding charge-density $\rho_s$, by selecting $K$ so that $q=0$, the $\he$-projection of $\gae$ is the scaled dipole-moment $\vp/D_0$.  Alternative methods to define $\gae$ exists see \eg~\cite{Schiffer+etal1949}. 

The electric polarizability tensor $\gae$ is a symmetric positive semi-definite tensor~\cite{Sjoberg2009b}. Note also that $\gae$ depend only on the domain $V$. For the sphere of radius $a$ we have $\rs=3E_0\he\cdot \hr$, with corresponding dipole moment $\vp=4\pi a^3 E_0\he$, and potential $\phi=E_0 a^3(\he\cdot \hr)/(\eps r^2)$, and thus $\gae=4\pi a^3 I$.

The electric polarizability tensor appears in the literature with several different notations, in~\cite{Schiffer+etal1949} it is denoted $e_{jk}$, in~\cite{Payne1967,Kleinman+Senior1986} a related quantity is called $P_{jk}$ and in~\cite{Jackson1999,Gustafsson+etal2009a} and subsequent publications it is denoted $\gamma$ or $\gae$ and in~\cite{deMeulenaere+vanBladel1977,Arvas+Harrington1983} it is denoted $\pi_{\mrm{e}}$ and $\pi_{\mrm{e}}^d$ respective, and~\cite{Sihvola2005,Yaghjian+etal2013} it is denoted $\alpha$. For generalization of $\gae$ to a larger class of materials as well as a review of its properties see \eg~\cite{Sjoberg2009b}.

\subsection{Magnetic polarizability tensor}

Magnetic polarizability is defined analogous to electric polarizability, as the map between a boundary condition and the behavior at infinity.  
Given an external field $B_0\hh$ we define a current density $\vJ$ and the associated vector potential $\vA$ that are divergence free. Formally, we do this by introducing an energy space $X$ of $\vJ$ such that $\int_V\int_V \vJ^*(\vr_1)\cdot\vJ(\vr_2)/|\vr_1-\vr_2|\diff V_1\diff V_2<\infty$, and subsequently define
\begin{equation}\label{X0}
X_0=\{\vJ\in X; \Div\vJ=0\}. 
\end{equation}
in the distributional sense. Current densities throughout this paper are in $X_0$ or subsets of $X_0$.

Our starting point is here to consider the current density $\vJ\in X_0$ that is the solution to the integral equation:
\begin{equation}\label{vJ}
\mu\int_V \frac{\vJ(\vr_2)}{4\pi|\vr_1-\vr_2|}\diff V_2 = \frac{1}{2} \mu H_0\hh \times \vr_1 , \ \vr_1\in V.
\end{equation}
The currents are due to the applied external magnetic field $\vH=H_0\hh=\frac{1}{\mu}B_0\hh$. Note that if we operate on both sides within the volume with the operator $\Rot\Rot$ we see that $\vJ$ only have support on the boundary, formally we use $\vJ\diff V=\Js\diff S$, similar to the charge density case above. For surface current densities $\Js$, we note that the divergence free condition~\eqref{X0} is equivalent with $\hn\cdot {\Js}=0$ and $\nabla_S\cdot \Js=0$, and we let $X_{0s}$ denote this subset of $X_0$. Here $\nabla_S\cdot $ is the surface divergence, \ie $\nabla=\hn\partial_n + \nabla_S$. The two degrees of freedom of the surface currents are determined by the equation:
\begin{equation}\label{scJ}
\hn \times \int_{\partial V} \frac{{\Js}(\vr_2)}{4\pi|\vr_1-\vr_2|}\diff S_2 = \frac{1}{2}\hn\times ( \hh H_0\times \vr_1), \ \vr_1\in \partial V.
\end{equation}
We recognize~\eqref{vJ} as a vector potential $\vA$ defined as
\begin{equation}
\frac{1}{\mu}\vA(\vr_1) = \int_{\partial V} \frac{\Js(\vr_2)}{4\pi|\vr_1-\vr_2|}\diff S_2, \
\vr_1\in \RR^3\backslash V,
\end{equation}
clearly $\vA$ is in the Coulomb gauge, \ie $\Div\vA=0$ in the exterior domain.
The multipole expansion of the vector-potential for $r\rightarrow \infty$ is given as
\begin{equation}\label{vA}
\frac{4\pi}{\mu}\vA(\vr_1) = \int_{\partial V} \frac{\Js(\vr_2)}{|\vr_1-\vr_2|}\diff S_2 \rightarrow \frac{1}{r^3_1}\int_{\partial V} (\vr_1\cdot\vr_2)\Js(\vr_2)\diff S_2 + \Oh(r^{-3}_1) = \frac{\vm\times \hr}{r^2_1}+\Oh(r^{-3}_1), \ \text{as}\ r_1\rightarrow \infty,
\end{equation}
where $\nabla_S\cdot{\Js}=0$ ensures that the magnetic monopole, $q_{\mrm{m}}/r$ term vanish. Here $\vm = \frac{1}{2}\int_{\partial V} \vr\times \Js\diff S$. The magnetic polarizability tensor $\gam$, as a $3\times 3$-matrix, is defined analogously to the electrical case:
\begin{equation}\label{magpol}
\gam \cdot \hh H_0 = \vm.
\end{equation}
We have here a choice of sign for $\gam$, the choice in~\eqref{magpol} ensures that $\gam$ is a positive semi-definite matrix for surface-currents in this case. Alternative sign-conventions exist in~\eqref{magpol} see \eg~\cite{Yaghjian+etal2013}. As an example for a sphere of radius $a$, we find that the surface current $\Js=(3/2)H_0 \hh \times \hr$ satisfy~\eqref{scJ}. The associated dipole-moment is $\vm=2\pi a^3\hh H_0$, and the magnetic polarizability is hence $\gam=2\pi a^3\bm{I}$, where $\bm{I}$ is a 3 times 3 unit tensor. A numerical approach to solve~\eqref{scJ} through the Method of Moments for the rectangular plate together with a singular value decomposition procedure to remove the gauge-freedom was used to determine the result depicted in Figure~\ref{fig:plate}. The induced magnetic moment $\vm$ for a fixed external $B_0\hh$, is large if we have a large current loop-area. Similarly to the electric polarizability measure of charge separation, we have here that the magnetic polarizability measure `current loop-area', orthogonal to the $\vB$-field direction. 

Given a permeable body in an external field, we note that the case considered above is when $\mu\rightarrow 0$ and the total field is given by $\vB=\vB_0 - \Rot \vA_p$, where $\vA_p=\vA$ as calculated above.

\subsection{Calculations of the magnetic polarizability tensor}

Given a smooth boundary $\partial V$, with a well-defined interior and exterior, we can similarly to the electric potential write a corresponding partial differential equation for the vector potential $\vA$ with a boundary condition $\mu H_0\hh$:
\begin{align}
\Rot\Rot \vA &= 0, \ \text{for} \ \vr\in \RR^3\backslash V,\\
\Div\vA &=0, \ \text{for} \ \vr\in \RR^3\backslash V,\\
\hn\times \vA &= \frac{1}{2}\hn\times (\mu H_0 \hh \times \vr),\ \vr\in \partial V,\\
\vA &\rightarrow \Oh(r^{-2}), \ \text{as}\ r\rightarrow \infty.
\end{align}
A fundamental solution approach of this vector Laplace-problem, \ie by expressing $\vA$ in terms of the single-layer operator yields the solution $\vA$ that satisfy~\eqref{vA}, where the surface current density ${\Js}$ is determined by solving the Fredholm integral equation~\eqref{vJ} of the first kind. 
Numerical approaches to solve this quasi-static problems for $\vA$ are given in~\eg~\cite{Bossavit1998,Albanese+Rubinacci1988,Lee+etal2003,Sanches+etal2010}, where a central piece is the conservation of the gauge-condition in the numerical basis element. 

However, for closed smooth non-degenerate domains there is an alternative approach to obtain $\gam$. Towards this end we note that the exterior domain is source free and we introduce the scalar magnetic potential $\pim$, with $\vH = \vH_0 + \nabla \pim$. Such a potential satisfy the Neumann problem for the scalar Laplace equation:
\begin{align}
\Laplace \pim &= 0, \ \text{for}\ \vr\in\RR^3\backslash V,\nonumber\\
-\partial_n\pim & = H_0\hh \cdot \hn \ \text{on}\ \vr\in \partial V,\label{N}\\
\pim &\rightarrow \Oh(r^{-2}) \ \text{as} \ r\rightarrow 0. \nonumber
\end{align}
A fundamental solution approach with an associated charge-density results in the relations:
\begin{equation}\label{Nsol}
4\pi\pim(\vr)=\int_{\partial V} \frac{\rms(\vr')}{|\vr-\vr'|}\diff S' \rightarrow \frac{q_m}{r} + \frac{\vm\cdot\hr}{r^2}+\Oh(r^{-3}), \ \text{as}\ r\rightarrow \infty.
\end{equation}
Note that the term $q_{\mrm{m}}$ vanish on closed surfaces due to that $\oint_{\partial V} \rms\diff S = H_0\int_V \Div\hh\diff V=0$. 
The magnetic charge density $\rms$ is determined through a Fredholm integral equation of the second kind~\cite{Jaswon+Symm1977,Folland1995}: 
\begin{equation}\label{Neumann}
 \frac{1}{2}\rms + \int_{\partial V} \frac{\hn\cdot(\vr-\vr')}{4\pi|\vr-\vr'|^3} \rms(\vr')\diff S' = H_0\hh\cdot\hn, \ \vr\in \partial V.
\end{equation}
The factor $1/2$ is associated with that the boundary is locally smooth, for a corner or line the corresponding volume-angle normalized with $4\pi$ appears.   
Here we have $\vm  = \int_{\partial V} \vr\rms\diff S$, the magnetic polarizability now follows from 
\begin{equation}
\gam\cdot \hh H_0 = \vm.
\end{equation}

If the volume $V$ is simply connected with a sufficiently smooth boundary, then \eqref{N} is uniquely solvable and the solution is given by \eqref{Nsol}, also for domains where the exterior have disconnected parts we have uniqueness, see~\eg~\cite{Folland1995} due to that $\hh$ is constant. If we return to the sphere, we note that $\rms=(3H_0/2)\hh\cdot\hr$ solves~\eqref{Neumann} with magnetic scalar potential $\pim=(H_0 a^3/2)\hh\cdot \hr/r^2$ that satisfy~\eqref{N}, and the associated magnetic dipole moment is $\vm=2\pi a^3 H_0 \hh$, and consequently we find again $\gam = 2\pi a^3 \bm{I}$, where $\bm{I}$ is a 3 times 3 unit tensor.

Scattering problems that connect a dipole moment to the magnetic field have been studied in~\cite{Rayleigh1871,Bethe1944} and with explicit polarizability tensor in \eg~\cite{Kleinman+Senior1986,deMeulenaere+vanBladel1977,Arvas+Harrington1983}, the context is analytic and numerical implementation to determine electric and magnetic dipole-moments of planar apertures. 

We note that $\gae$ and $\gam$ corresponds to solving the scalar Dirichlet and Neumann problem respectively for the scalar Laplacian in an exterior domain. There are a few different normalizations and sign-conventions of $\gam$, in~\cite{Schiffer+etal1949} their corresponding dipole-form $d_{jk}=4\pi (\gam)_{jk}$. Another normalization for small surfaces $S$, $\gam/|S|^{3/2}$, is used in \eg~\cite{deMeulenaere+vanBladel1977} to make the quantity independent of equivalent volume, here $|S|$ is the area of $S$, see also~\cite{Payne1967,Yaghjian+etal2013,Sjoberg2009b}, for additional properties and different sign-conventions.

\section{Alternative derivation of $\rP$ for the electrically small case}\label{app:long}

An alternative method to determine the leading order behavior of $\rP$, given in~\eqref{smallrP}, for the electrically small case, $ka\ll 1$, is to start directly from \eqref{oprP} and using the EFIE operators $\LL=\LL_e-\LL_m$ defined in~\eqref{Le} and \eqref{Lm}.  The explicit expression, using these definitions is:
\begin{multline}\label{Prad}
\rP  
= \imp\iint_{V\times V} \Big[\big(k^2 \Jeo \cdot \Jet^*
-(\Div\Jeo)(\Div \Jet^*)\big) + \\ 
\frac{1}{\imp^2}\big(k^2 \Jmo\cdot \Jmt^*-(\Div\Jmo) (\Div \Jmt^*) \big)\Big]\frac{\sin(kR)}{8\pi k R}\diff V_1\diff V_2 
 \\+ \frac{k^2}{4\pi}\iint_{V\times V}   \sbj_1(kR)\hat{\vR} \cdot \IM (\Jeo^*\times\Jmt) \diff V_1\diff V_2,
\end{multline}
where $\Jeo=\vJ_{\mrm{e}}(\vr_1)$, $\Jet=\vJ_{\mrm{e}}(\vr_2)$ and analogously for $\Jmo$ and $\Jmt$, as usual $\vR=\vr_1-\vr_2$, $R=|\vR|$ and $\hR=\vR/R$.  
We expand the integrand in terms of small $ka$ the dependence of the current-densities on $k$ is accounted for by inserting the current expansion~\eqref{exp} into~\eqref{Prad}.

Note that the integrand consists of terms, $\vJe$, $\vJm$ and mixed terms. The pure $\vJe$ and the pure $\vJm$-terms are equal in structure (up to the constant $\eta^2$). For the integrand with purely electrical terms in~\eqref{Prad} we find by inserting~\eqref{exp}:
\begin{multline}\label{C2}
\frac{\imp k^2}{8\pi}\Big\{\Jeo^{(0)}\cdot\Jet^{(0)*} - (\Div\Jeo^{(1)})(\Div\Jet^{(1)*})
 + 2k \RE\big[\Jeo^{(0)}\cdot\Jet^{(1)*} - (\Div\Jeo^{(1)})(\Div\Jet^{(2)*})\big]  \\ 
 +k^2
\Big(\Jeo^{(1)}\cdot\Jet^{(1)*} - (\Div\Jeo^{(2)})(\Div\Jet^{(2)*}) 
+ 2\RE \big(\Jeo^{(0)}\cdot\Jet^{(2)*} - (\Div\Jeo^{(1)})(\Div\Jet^{(3)*}) \big]\Big)\\+\Oh(k^3)\Big\}
\big[1 -  \frac{(kR)^2}{6}+\Oh(k^3)\big].
\end{multline}
We recall that \eqref{C2} is part of the integrand in~\eqref{Prad}, we note that upon integration several of the above terms vanish by using~\eqref{zero} and \eqref{zeroone}. 
The first electrical non-vanishing contribution term in the integrand to $\rP$ are of $k^4$-order and have an integrand of the form:
\begin{equation}\label{Prad-survivors}
\frac{\imp k^4}{8\pi}\Big\{
\Jeo^{(1)}\cdot\Jet^{(1)*} 
 +
\frac{ \vr_1\cdot\vr_2}{3}\big[\Jeo^{(0)}\cdot\Jet^{(0)*} - (\Div\Jeo^{(1)})(\Div\Jet^{(1)*})
\big]\Big\}
+\Oh(k^5).
\end{equation}
The pure magnetic terms give an analogous expression to~\eqref{Prad-survivors}.

For the cross term, $\Jet^*\times\Jmt$ in \eqref{Prad}, we first note that $\sbj_1(kR)=(kR/3)[1-(kR)^2/10+\Oh(k^4)]$, and that we recall $R\hR=\vR$, to find the integrand
\begin{equation}
\frac{k^3}{12\pi}\big[1-\frac{(kR)^2}{10}+\Oh(k^4)\big]\vR\cdot 
\IM(\Jeo^{(0)*}\times \Jmt^{(0)} 
 +k(\Jeo^{(0)*}\times \Jmt^{(1)} +\Jeo^{(1)*}\times \Jmt^{(0)}) +\Oh(k^2) ).
\end{equation}
Through~\eqref{zero}, we find that the apparent leading order $\int_V\int_V \vR\cdot\Jeo^{(0)*}\times\Jmt^{(0)} \diff V_1\diff V_2$ vanish and the $k^4$-order terms is the first remaining term. Putting all these results together yields the radiated power:
\begin{multline}\label{arp}
\rP =  
 \frac{k^4}{4\pi}\Big[ \frac{\imp}{2}\Big\{ |\int_V \vJe^{(1)}\diff V|^2  
 + \frac{1}{3}\int_V \int_V \vr_1\cdot\vr_2\big[\Jeo^{(0)}\cdot\Jet^{(0)*} -(\Div \Jeo^{(1)})(\Div \Jet^{(1)*})\big]\diff V_1\diff V_2\Big\} \\ 
+ \frac{1}{2\imp}\Big\{|\int_V \vJm^{(1)}\diff V|^2  
 + \frac{1}{3}\int_V \int_V \vr_1\cdot\vr_2\big[\Jmo^{(0)}\cdot\Jmt^{(0)*} - (\Div \Jmo^{(1)})(\Div \Jmt^{(1)*})\big]\diff V_1\diff V_2 
\Big\}
\\+ \frac{1}{3} \IM\Big\{
\int_V \vJm^{(1)}\diff V\cdot \int_V \vr\times \vJe^{(0)*}\diff V +
\int_V \vJe^{(1)*}\diff V\cdot \int_V \vr\times \vJm^{(0)}\diff V\Big\} 
\Big] + \Oh(k^5).
\end{multline}
Partial integration yields $\int_V \vJe^{(1)}\diff V = \ju c\vpe$, where $c=1/\sqrt{\eps\mu}$ is the speed of light. Similar vector manipulation~\cite[p433]{Stratton1941}~\cite[p127]{Kristensson1999a} yields $\int_V (\hr_2\cdot \vr_1)\Jeo^{(0)}\diff V_1=\vme\times \hr_2$ since $\Div\vJe^{(0)}=0$. Here $\vpe=\int_V \vr\roe^{(1)}\diff V$, where $\roe^{(1)}=-\Div\vJe^{(1)}/c$, and $\vme=\frac{1}{2}\int_V \vr\times \vJe^{(0)}\diff V$, and analogously for the magnetic currents. Inserting these expressions reduce the total radiated power, \eqref{arp}, to the contribution from the electric current dipoles $(\vpe,\vme)$ and the magnetic current dipoles $(\vpm,\vmm)$ as
\begin{equation}
\rP =  
 \frac{k^4}{12 \pi \sqrt{\eps\mu}}\Big[ \big|\frac{1}{\sqrt{\eps}}\vpe - \sqrt{\eps}\vmm\big|^2  
   + \big|\frac{1}{\sqrt{\mu}}\vpm +\sqrt{\mu}\vme\big|^2 \Big] 
+\Oh(k^5) .
\end{equation}
An equivalent, but for optimization more tractable expression given current densities is:
\begin{equation}\label{rpsmall}
\rP =  
 \frac{k^4\imp}{12 \pi }\Big[ \big|\int_V \ju\vJe^{(1)} + \frac{1}{2\eta} \vr\times \vJm^{(0)} \diff V \big|^2  
   + \big|\int_V \frac{\ju}{\eta}\vJm^{(1)} -\frac{1}{2}\vr\times \vJe^{(0)}\diff V\big|^2 \Big] 
+\Oh(k^5) ,
\end{equation}
which is identical to~\eqref{smallrP}.

\section{Polarizability tensors for an ellipsoidal shape}\label{app:ell}

For ellipsoidal shapes we follow \eg~\cite{Schiffer+etal1949,Sihvola2005,Gustafsson+etal2007a} and define the dimensionless quantity of the depolarization factors
\begin{equation}
L_j=\frac{a_1a_2a_3}{2} \int_0^\infty \frac{\diff s}{(s+a_j^2)\sqrt{(s+a_1^2)(s+a_2^2)(s+a_3^2)}},
\end{equation}
where $a_1,a_2,a_3$ denote the radii of the three axis.

The electric and magnetic polarizability tensors are
\begin{equation}\label{egem}
	(\gae)_{jj}=\frac{V}{L_j} \ \text{and} \ (\gam)_{jj}=\frac{V}{1-L_j},\ \text{for}\ j=1,2,3
\end{equation} 
all other elements in $\gae$ and $\gam$ are zero given that the coordinate axis are co-oriented and centered with the axis of the ellipsoidal. Here the volume is $V=4\pi a_1a_2a_3/3$.

We consider the cases of a flat ellipse, an ellipsoidal with a small $a_3$, and the oblate and prolate spheroidal. We note that the integral above is given in terms of elliptic incomplete integrals as:
\begin{align}
L_1 &= \frac{\zeta\xi}{\sqrt{1-\zeta^2}(1-\xi^2)}(\Fint(\arccos(\zeta),\frac{1-\xi^2}{1-\zeta^2})-\Eint(\arccos(\zeta),\frac{1-\xi^2}{1-\zeta^2}),\label{D2}\\
L_2 &= \zeta\frac{-(1-\xi^2)\zeta+\xi\sqrt{1-\xi^2}\Eint(\arccos(\xi),\frac{1-\zeta^2}{1-\xi^2})}{(1-\xi^2)(\xi^2-\zeta^2)},\\
L_3 &= \xi\frac{-\sqrt{1-\zeta^2}\xi+\zeta \Eint(\arccos(\zeta),\frac{1-\xi^2}{1-\zeta^2})}{\sqrt{1-\zeta^2}(\zeta^2-\xi^2)},
\end{align}
where we have used $a_1=a,a_2=\xi a, a_3=\zeta a$ with $0<\xi<1$ and $0<\zeta<1$ and the identity $\Eint(\iu \phi,k^2)=\iu(\Fint(\psi,(k')^2)-\Eint(\psi,(k')^2)+\tan\psi\sqrt{1-k'^2\sin^2\psi})$, where $\sinh\phi=\tan\psi$, $k'=\sqrt{1-k^2}$~\cite[19.7.7]{dlmf.nist.gov} to simplify $L_1$ to \eqref{D2}. The following notation is used for the incomplete elliptic integrals of first and second kind: $\Fint(\alpha,m)=\int_0^{\alpha} \diff \theta/\sqrt{1-m\sin^2\theta}$, $\Eint(\alpha,m) = \int_0^\alpha \sqrt{1-m\sin^2\theta}\diff \theta$, and $\Kint(m)=\Fint(\pi/2,m)$ and $\Eint(m)=\Eint(\pi/2,m)$ for the complete elliptic integrals of first and second kind~\cite{dlmf.nist.gov}. 

For the case with zero-thickness, or $a_3=a\zeta\rightarrow 0$, and $a_1=a$, $a_2=a\xi$, $0<\xi<1$ we find that the depolarization factors reduce to:
\begin{equation}
\frac{L_1}{\zeta}=\xi\frac{\Kint(1-\xi^2)-\Eint(1-\xi^2)}{1-\xi^2}, \
\frac{L_2}{\zeta}=\frac{\Eint(1-\frac{1}{\xi^2})-\Kint(1-\frac{1}{\xi^2})}{1-\xi^2},\  
\frac{1-L_3}{\zeta} = \frac{\Eint(1-\xi^2)}{\xi}.
\end{equation}
As $\xi\rightarrow 0$, \ie the needle, we find that 
\begin{equation}\label{d5}
\frac{L_1}{\zeta}=\xi(\log(\frac{4}{\xi})-1)+\Oh(\xi^3),\
\frac{L_2}{\zeta}=\frac{1}{\xi}+\Oh(\xi),\ 
\frac{1-L_3}{\zeta} = \frac{1}{\xi}+\Oh(\xi)
\end{equation}
and as $\xi\rightarrow 1$, \ie the disc, we find that 
\begin{equation}\label{d6}
\frac{L_1}{\zeta} = \frac{\pi}{4}+\frac{1}{16}\pi(\xi-1)+\Oh(\xi-1)^2,\
\frac{L_2}{\zeta} = \frac{\pi}{4}-\frac{5}{16}\pi(\xi-1)+\Oh(\xi-1)^2,\
\frac{1-L_3}{\zeta} =\frac{\pi}{2}+\frac{\pi}{4}(1-\xi)+\Oh(1-\xi)^2.
\end{equation}

The polarizabilities for the case of $\zeta\rightarrow 0$ are 
\begin{equation} 
(\gae)_{11} = \frac{4\pi a^3 \zeta \xi}{3 L_1}, \ \ 
(\gae)_{22}= \frac{4\pi a^3 \zeta\xi}{3L_2}, \ \ 
(\gam)_{33} = \frac{4\pi a^3 \zeta\xi}{3(1-L_3)}.
 \end{equation}
All other elements of $\gae$, $\gam$ vanish for $\zeta\rightarrow 0$, see~\eqref{egem}, the shape is reflection symmetric and hence is both $\gae$ and $\gam$ diagonal in a coordinate-system where the reflection symmetries coincide with the coordinate axes, see~\cite{Kleinman+Senior1986}. Note that as $\xi\rightarrow 1$, we recover the known value of the disc with $(\gae)_{11}=16a^3/3$ and $(\gam)_{33}=8a^3/3$. These eigenvalues and their corresponding antenna Q factors are depicted in Figure~\ref{fig:flatE}.

For a rotationally symmetric ellipsoid, \ie the spheroidal shape, we have two cases, the oblate and the prolate case. For the oblate case $a_1=a_2=a,a_3=a\zeta$ and we find
\begin{align}
L_1=L_2=\frac{\zeta}{2(1-\zeta^2)^{3/2}}\big(\arccos(\zeta)-\zeta\sqrt{1-\zeta^2}\big)\\
L_3=\frac{1}{(1-\zeta^2)^{3/2}}\big(\sqrt{1-\zeta^2}-\zeta\arccos(\zeta)\big).
\end{align}
For the prolate case $a_1=a_2=\zeta a,a_3=a$
\begin{align}
L_1=L_2=\frac{1}{2(1-\zeta^2)^{3/2}}\Big(\sqrt{1-\zeta^2}-\zeta^2\ln(\frac{\zeta}{1-\sqrt{1-\zeta^2}})\Big),\\
L_3=\frac{\zeta^2}{(1-\zeta^2)^{3/2}}\Big( \ln(\frac{\zeta}{1-\sqrt{1-\zeta^2}}-\sqrt{1-\zeta^2}) \Big) ,
\end{align}
which agree with \eg~\cite{Gustafsson+etal2007a}, upon using standard identities. 
The corresponding eigenvalues and antenna Q are shown in Figure~\ref{fig:gam}.
An alternative approach is also given in~\cite{Kleinman+Senior1986}.

\section*{Acknowledgements}
We gratefully acknowledge the support from Swedish Governmental Agency for Innovation Systems (Vinnova), The Swedish Foundation for Strategic Research (SSF) and The Swedish Research Council (VR). 



\end{document}